\documentclass[pre,aps,twocolumn,showpacs]{revtex4}
\usepackage{graphicx}
\usepackage{amsmath}
\usepackage{amssymb}
\def\slaninafigdir{.}
\begin{document}
\title{%
Interacting molecular motors: Efficiency and work fluctuations
}%%
\author{%
Franti\v{s}ek Slanina
}
\affiliation{%
Institute of Physics,
 Academy of Sciences of the Czech Republic,
 Na~Slovance~2, CZ-18221~Praha,
Czech Republic
}
\email{
slanina@fzu.cz
}
\begin{abstract}
We investigate the model of ``reversible ratchet'' with interacting
particles, introduced by us earlier  [Europhys. Lett.
 {\bf 84},
 50009
 (2008)]. We further clarify the effect of efficiency enhancement due
to interaction and show that it is of energetic origin, 
rather than a consequence of reduced fluctuations. 
We also show complicated structures emerging in the
interaction and density dependence of the current and response
function. The fluctuation properties of the work and input energy
indicate in detail the far-from-equilibrium nature of the dynamics.
\end{abstract}
\date{\today}%
\maketitle%
\section{Introduction}
Molecular motors 
\cite{schliwa_03,jul_adj_pro_97,astumian_97,rei_han_02,reimann_02,sch_woe_03,han_mar_nor_04,lip_cha_klu_lie_mul_06,kol_fis_07,wan_els_07} 
are subject to intense study both from biological and
technological point of view. They are paradigmatic examples of
machines operating at nanometer scale. 
In a cell, motor proteins powered by ATP hydrolysis 
\cite{svo_blo_94,wan_ost_98,as_bi_96,car_cro_05}
help move molecules to places where they
are needed. Motors assist separation of chromosomes, copying  DNA
into RNA and perform many more functions
\cite{sch_xia_mer_wei_97,ras_kob_mal_fid_mas_04,del_benz_slu_aze_gol_06}.  
Technological applications of the underlying mechanisms flourish
\cite{han_mar_08},  including e. g. Brownian pumps 
\cite{mat_mul_03,ket_rei_han_mul_00}
and quantum tunneling ratchets
\cite{lin_hum_lof_99}.
They provide also an
invaluable testing ground for fundamental questions of transport
phenomena far from equilibrium \cite{bly_eva_07}. 

Many models of molecular motors appeared in the literature since the
pioneering work by Ajdari and Prost \cite{ajd_pro_92}. The basic
mechanism is best elucidated in the models which
rely on the ratchet mechanism
\cite{magnasco_93,bar_han_kis_94,kol_wid_98,rei_han_02,reimann_02} and
also bear the name Brownian motors. The basic idea can be viewed
either as diffusive
motion of a particle in spatially asymmetric time-dependent potential
or as chemically driven transitions between a finite number of
mechano-chemical states. The former view is more intuitive, but the
latter is closer to reality and opens the perspective of fitting the
underlying transition probabilities to experimental data.
 
More realistic models are rather built on Markov chains in the
configuration space constructed as product of spatial and internal
(chemical) coordinates \cite{qian_97,qian_00a,qian_04,mae_wie_03,kol_stu_pop_05,wan_fen_zhe_fan_07,lip_cha_klu_lie_mul_06,lip_lie_08,das_kol_08,vilfan_09}.
This approach resides perhaps on more solid experimental
evidence, but the absence of explicit potential makes it very
difficult to assess the energetic efficiency, the question of
principal importance in this paper.

Indeed, one of the points of special interest here will be the question of the
efficiency of molecular motors. Several measures of efficiency can be
found in literature. We shall use the classical thermodynamic
definition $\eta=W/E_\mathrm{in}$, where $W$ is work
performed and $E_\mathrm{in}$ energy  supplied to the system from
external source. Alternative measure takes into account viscous resistance
 \cite{wan_ost_02}, thus reflecting better the reality, at the cost
 that the inequality $\eta<1$ is not guaranteed automatically. Yet
 other methods of measuring the efficiency involve explicitly 
 the consumption of chemical energy \cite{jul_pro_95}, or the
magnitude of the stopping  force \cite{kol_fis_07}. Note, however,
that the former work (\cite{jul_pro_95}) explores the interacting
motors and the mechanism if generating the non-zero current is
related to spontaneous symmetry breaking and this it is 
principally different from the non-interacting case studied in
\cite{kol_fis_07}.
 Therefore, the
direct comparison of the efficiency in these two cases is hardly
possible. We are not aware of any work in which several measures of
efficiency would be systematically compared on the same model.

The efficiency of canonical
Brownian motors realized as either flashing or rocking ratchets was
intensely studied 
\cite{sekimoto_96,sekimoto_97,kam_hon_tak_98,par_dec_02,qian_04,asf_bek_05,par_bla_cao_bri_98,qian_97,qian_00a}.
It turns out that the energetic efficiency is rather low
\cite{reimann_02,par_dec_02}, while  the 
experimental data on motor proteins, e. g. the kinesin
\cite{svo_blo_94,kol_fis_07}, report high efficiency, sometimes even
estimated to be close to $100$ per cent. We are not in a position to
judge the quantitative precision of these empirical estimates,
although it can be suspected that the error margin is rather
high.
 However,
one is lead to a natural conclusion that the usual ratchet mechanism
with diffusion as principal driving force is not an appropriate model
for biological motors.

In idealized case we can distinguish between ratchet and power-stroke
mechanisms for molecular motors \cite{wan_ost_02a},
the latter relying rather on quasi-deterministic downhill motion in
a free-energy landscape which evolves in time. Thus, the particles
move as if trapped in a traveling potential wave. This idea was
elaborated in a toy model of ``reversible ratchet''
\cite{par_bla_cao_bri_98,parrondo_98,ast_der_99}, showing much higher
efficiency, close to the biologically relevant figures. Of course,
arbitrary combinations and mixtures of the ratchet and power-stroke
mechanisms can be invented and indeed, they are believed to be found
in reality, e. g. in the myosin V motor (see the review
\cite{vilfan_09} and references therein). 
Nevertheless, it is useful to compare
these two extremes. We should also
note that high
efficiency was  characteristic of the models of
either interacting \cite{jul_pro_95} or non-interacting
\cite{par_jul_ajd_pro_99,sch_sei_08} motors, which do combine the ratchet and
power-stroke mechanisms. 

The second point we shall concentrate on in this work will be the mutual
repulsive interaction of molecular motors. In the cell, the steric
(hard-core) repulsion of motor proteins influences significantly their
behavior. For example, in gene transcription and  
translation large number of motor proteins move along the same track
\cite{sch_xia_mer_wei_97,ras_kob_mal_fid_mas_04}, forming the so-called
``Christmas tree'' structures. Thus, they show themselves as  a
physical realization 
of the well-studied asymmetric exclusion process, introduced first in
the context of biopolymerization \cite{mcd_gib_pip_68,mcd_gib_69} and
later solved exactly in great detail, using sophisticated techniques
\cite{bly_eva_07,do_do_mu_92,de_ev_pa_93,derrida_98}. The model was
adapted for molecular motors, which may attach and detach with defined
rates \cite{par_fra_fre_03,par_fra_fre_04}.
Later, this situation was studied theoretically for the cases of  kinesin
\cite{gre_gar_nis_sch_cho_07}, ribosomes  \cite{bas_cho_07}, and RNA
polymerase \cite{tri_cho_08}, using the procedures developed in
traffic models \cite{na_sch_92}.  The influence of the geometry of the
compartment in which the interacting motors diffuse after detachment from the track
was studied e. g. in \cite{lip_klu_nie_01}.

Interaction of motors brings about even more complicated collective
effects. In the cell, kinesin and dynein
molecules 
typically carry the cargo in groups
\cite{lip_cha_klu_lie_mul_06,klu_lip_05,kol_fis_07}, resulting in
current reversals. 
Including explicitly the ratchet
 mechanism of  driven diffusion of hard-rod particles  leads to very
 intricate effects \cite{der_vic_95,der_ajd_96,agh_men_pli_99}, if the
 particle size and the ratchet periodicity are incommensurate. 
The collective movement of coupled Brownian motors was studied
\cite{rei_kaw_bro_han_99,stu_phi_kol_05,stu_kol_06,cqm_kqf_zel_cas_joa_06} 
and in some cases the coupling was found to induce
non-zero current and spontaneous oscillations 
even in mirror symmetric potential due to dynamical
symmetry breaking \cite{jul_pro_95,jul_pro_97}. In analogy with these
 works, the motion of a few
 rigidly bound motors was studied \cite{bad_jul_pro_02}. A special
 case of such interaction is the coordination of
 the two motor heads within single kinesin molecule, which leads also to
 non-trivial effects \cite{das_kol_08}. Finally, let us mention  the 
interaction of the motors with the track, studied in 
 ``burnt-bridge'' models, e. g. in Ref. 
\cite{art_mor_kol_08}.

\begin{figure}[t]
\begin{center}
\includegraphics[scale=0.5]{%
\slaninafigdir/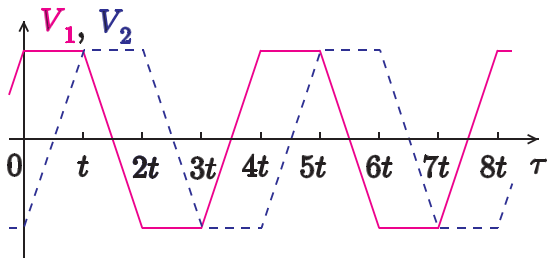}
\end{center}
\caption{%time-sequence....
(Color online) 
Graph of the time dependence of the potential in which the particles
move. Full line: $V_1(\tau)$, dashed line: $V_2(\tau)$.
 The potential at the third site, $V_0$, is time independent.
} 
\label{fig:time-sequence}
\end{figure}

In our previous paper \cite{slanina_08} we introduced a model, which
is a  modified
version of the ``reversible ratchet''. Spatial coordinate is
discretized, as e. g. in \cite{kol_wid_98}. Tunable on-site repulsion
between particles is introduced. We found  in \cite{slanina_08} that
not too strong  
interaction leads to increase of efficiency. This effect was
reproduced qualitatively in analytical calculations based on
mean-field (MF) approximation. Quantitative agreement was reached in an
improved MF treatment, developed in \cite{slanina_09}. 
Here we
investigate the model in depth by further numerical
simulations. Especially, we elucidate the origins 
of the interaction-enhanced efficiency. We shall show that it 
stems from the energy balance rather than from suppression of fluctuations.
 At stronger interaction and/or higher
density, current reversals and oscillations in response function are
found. We also calculate the distribution of input
energy and performed work, which is far from being Gaussian.

\begin{figure*}[t]
~~~~~~~~~~~~~~~~~~~~~~~~~~~~~$N=9$\hfill$N=12$\hfill$N=18$~~~~~~~~~~~~~~~~~~~~~~~~\\ 
\includegraphics[scale=0.65]{%
\slaninafigdir/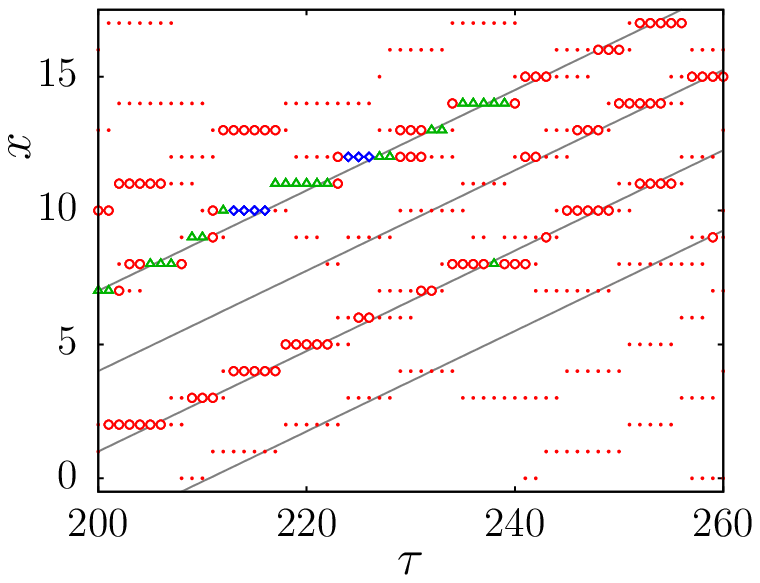}%
\includegraphics[scale=0.65]{%
\slaninafigdir/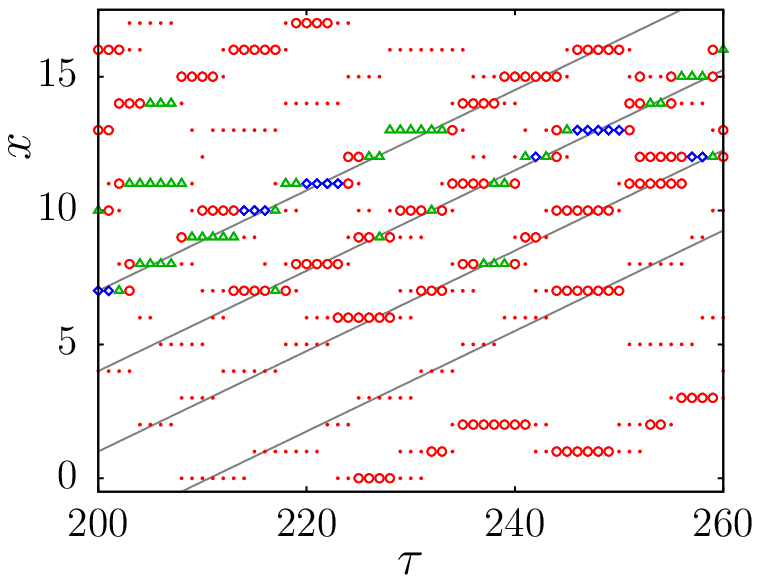}%
\includegraphics[scale=0.65]{%
\slaninafigdir/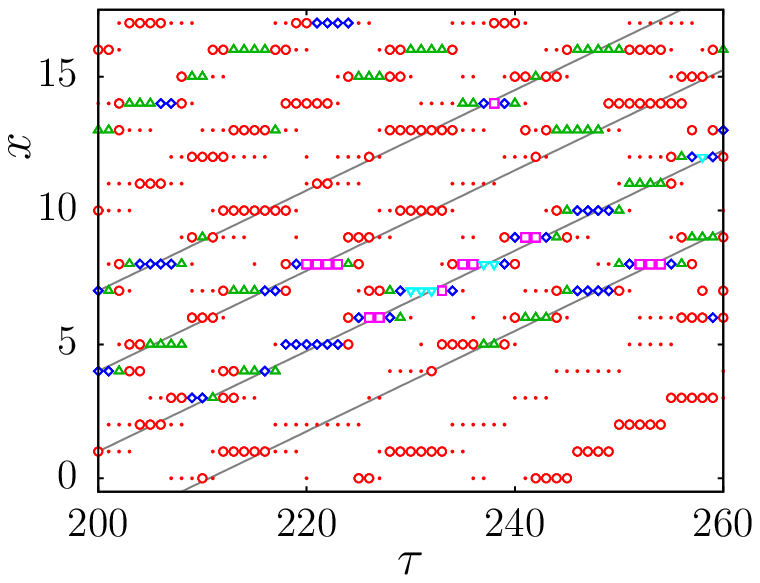}%
\rotatebox{90}{~~~~~~~~~~~~~~~~$g=0$}
\\
\includegraphics[scale=0.65]{%
\slaninafigdir/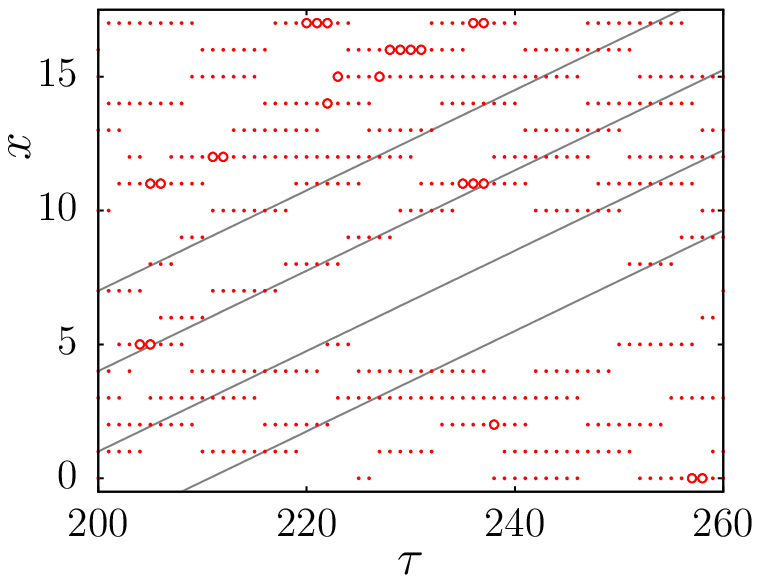}%
\includegraphics[scale=0.65]{%
\slaninafigdir/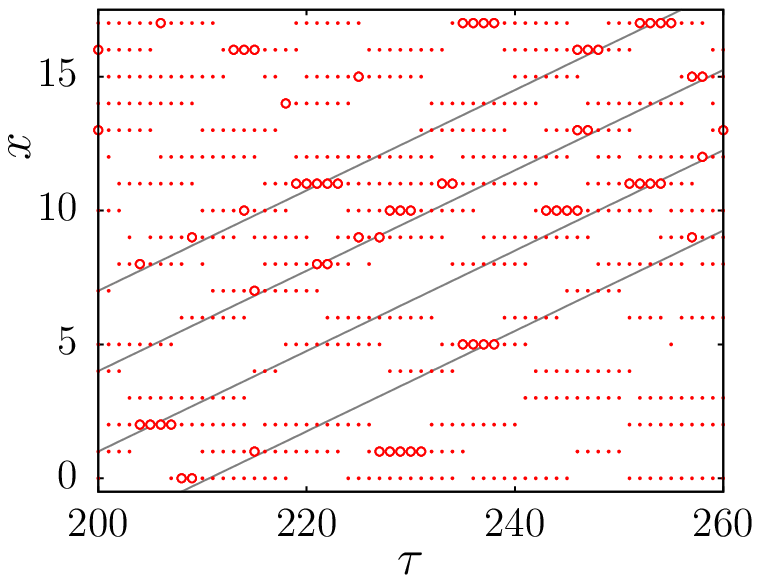}%
\includegraphics[scale=0.65]{%
\slaninafigdir/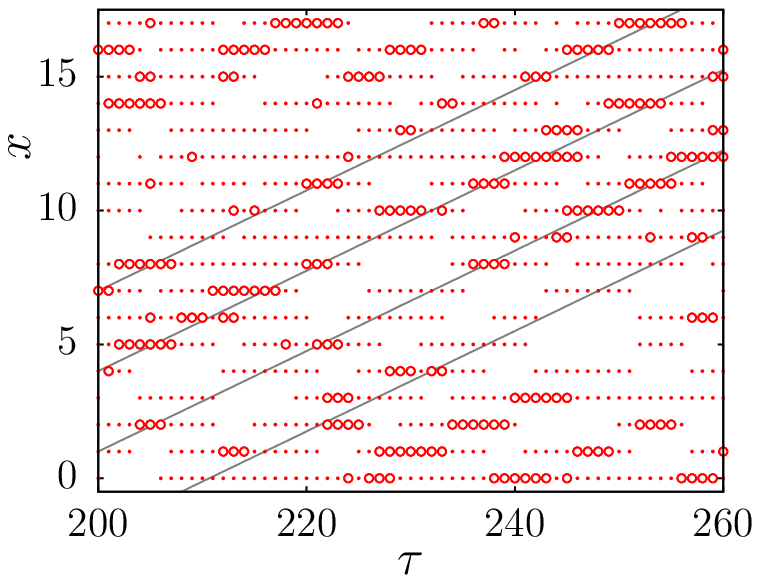}%
\rotatebox{90}{~~~~~~~~~~~~~~~~$g=0.5$}
\\
\includegraphics[scale=0.65]{%
\slaninafigdir/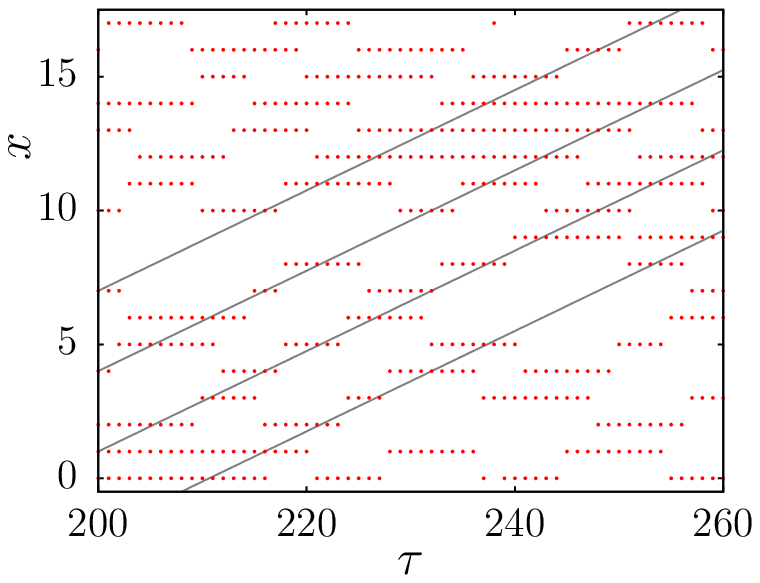}%
\includegraphics[scale=0.65]{%
\slaninafigdir/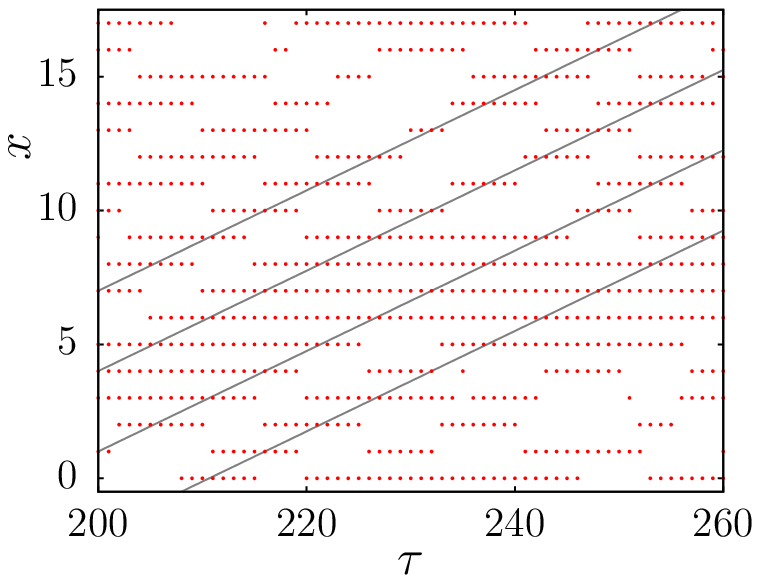}%
\includegraphics[scale=0.65]{%
\slaninafigdir/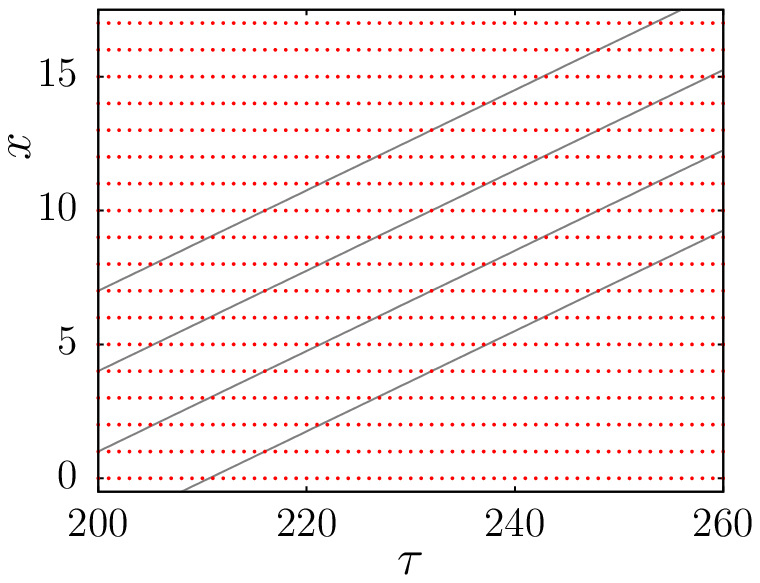}%
\rotatebox{90}{~~~~~~~~~~~~~~~~$g=0.9$}
\caption{%...
(Color online) 
Spatio-temporal diagrams of the configurations of the motor
particles. The width of the sample is $L=18$, temperature $T=10$,
quarter-period $t=4$, external load $F=0$. Each panel corresponds to
different combination of two parameters, the number of particles $N$
and the interaction strength $g$, whose values are indicated at
corresponding columns and rows. Dots denote presence of exactly one
particle at given space and time, 
the other symbols presence of more particles, namely two
({\Large $\circ$}), three ($\bigtriangleup$), four ($\diamond$), five
($\Box$), and six ($\bigtriangledown$).
The diagonal straight lines are guides for the eye, indicating the
movement of the minima of the potential $V(x,\tau)$.
} 
\label{fig:spacetime}
\end{figure*}

\begin{figure}[t]
\includegraphics[scale=0.85]{%
\slaninafigdir/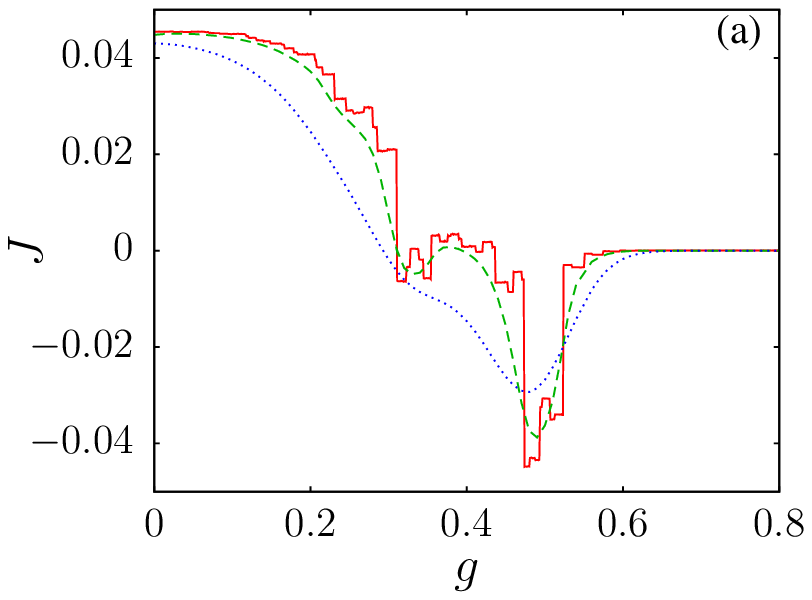}
\\
\includegraphics[scale=0.85]{%
\slaninafigdir/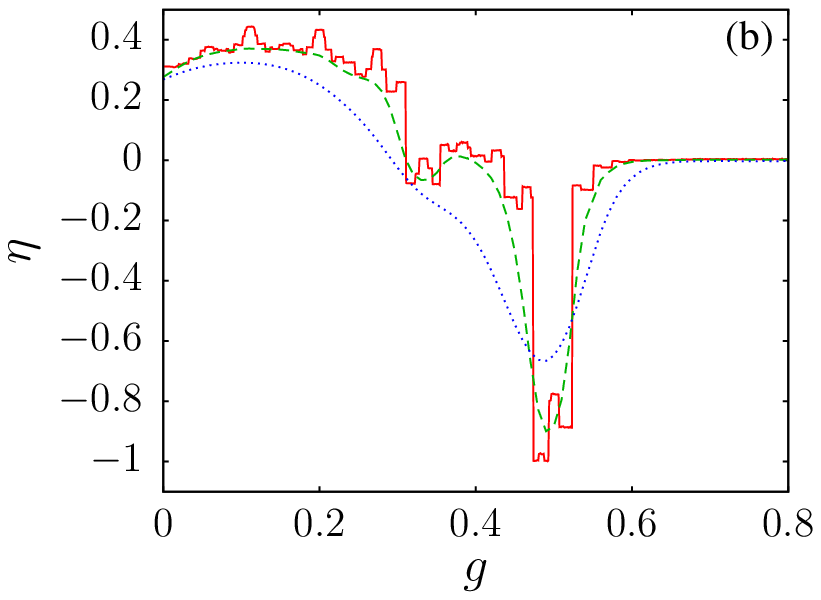}
\caption{%cur-vs-int-1200-1200-all-16-x-1...
%eff-vs-int-1200-1200-all-16-x-1...
(Color online) 
Current (panel (a)) and efficiency (panel (b)) as a function of
interaction strength, for $N=L=1200$, $t=16$, and $F=0.1$. The
temperature is $T=0$ (solid line), $T=10$ (dashed line), and $T=30$
(dotted line).}
\label{fig:eff-vs-int-1200-1200-all-16-x-1}
\end{figure}

\begin{figure}[t]
\includegraphics[scale=0.85]{%
\slaninafigdir/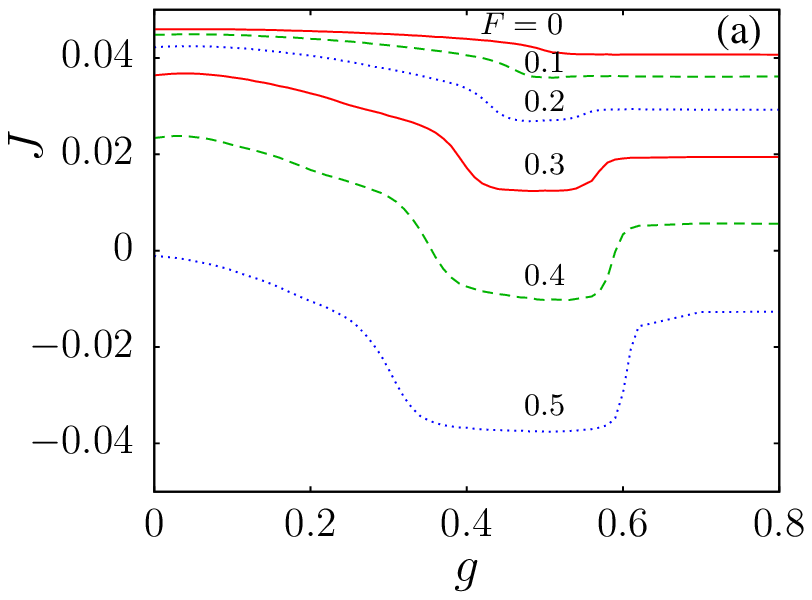}
\includegraphics[scale=0.85]{%
\slaninafigdir/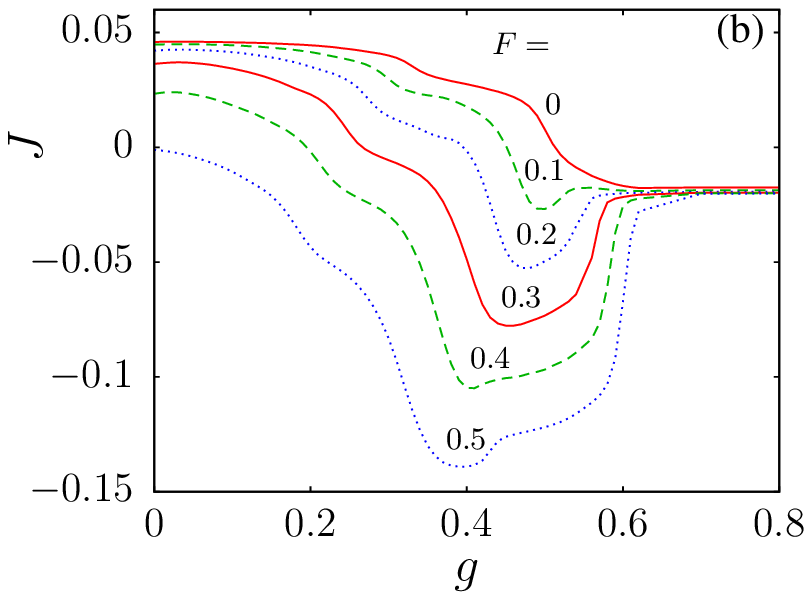}
\includegraphics[scale=0.85]{%
\slaninafigdir/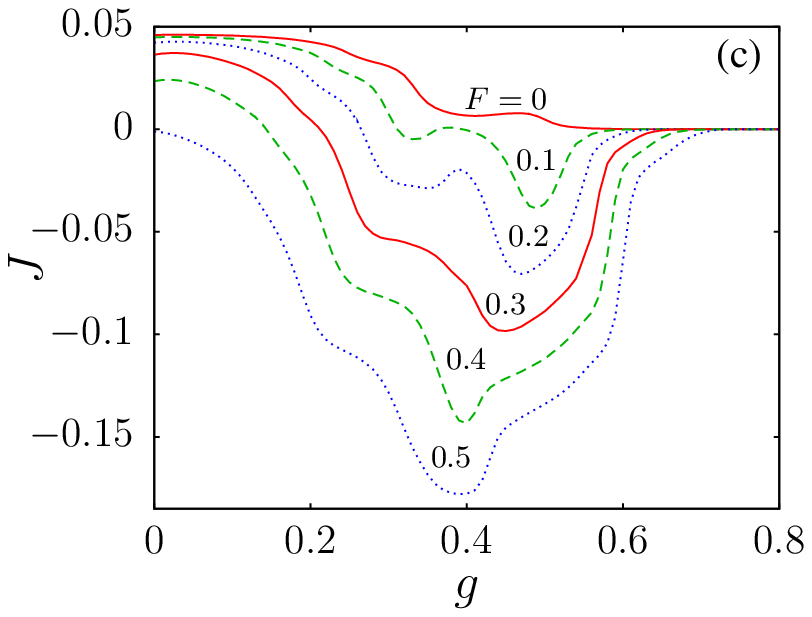}
\caption{%cur-vs-int-1200-360-10-16-x-all...
%cur-vs-int-1200-840-10-16-x-all...
%cur-vs-int-1200-1200-10-16-x-all...
(Color online) 
Current  as a function of
interaction strength, for $L=1200$, $t=16$, and $T=10$. The
number of particles is $N=360$ (density $\rho=0.3$) in the 
panel (a), $N=840$ (density $\rho=0.7$) in the  panel (b), and $N=1200$
(density $\rho=1$) in the panel (c). Different curves correspond to
load
, from top to bottom, $F=0$ (solid), $0.1$ (dashed), $0.2$ (dotted), $0.3$
(solid), $0.4$ (dashed), $0.5$ (dotted). 
}
\label{fig:cur-vs-int-1200-360-10-16-x-all}
\end{figure}

\section{Reversible ratchet with interacting particles}

Our model contains $N$ particles occupying integer positions on a
segment of length $L$, with periodic boundary conditions, and evolves
in discrete time.  The position of $i$-th particle at the instant
$\tau$ is denoted $x_i(\tau)$. The
particles move under the influence of a variable driving force with spatial
period $3$ and temporal period $4t$. The potential of this force is
$V(x,\tau)=V(x,\tau+4t)\equiv V_{x\mathrm{~mod~}3}(\tau)$, 
at site $x$ and time $\tau$.  The three independent values of the
potential $V_a(\tau)$, $a=0,1,2$ evolve in a four-stroke pattern, with
$V_0(\tau)=0$ and the remaining two being delayed one with respect to
the other by one quarter-period $t$. Thus, we prescribe
\begin{equation}
\begin{split}
V_1&(\tau)=V_2(\tau+t)=\\
=&\,\left\{
\begin{array}{rlrrrl}
&V&\mathrm{~for~}&0&<\tau<&t\\
&V+2V\,(1-\tau/t)&\mathrm{~for~}&t&<\tau<&2t\\
-&V&\mathrm{~for~}&2t&<\tau<&3t\\
-&V-2V\,(1-\tau/t)&\mathrm{~for~}&3t&<\tau<&4t\;.
\end{array}
\right.
\end{split}
\label{eq:time-dep-potential}
\end{equation}
We easily recognize the traveling-wave character of this potential,
corresponding to the power-stroke mechanism of the molecular-motor
movement. In all the rest of this paper, we fix the amplitude of the
potential $V=1$. The time dependence of the potential is illustrated
in Fig. \ref{fig:time-sequence}.

Besides the driving potential, there is also a uniform external force
from the useful load $F$ and, most importantly, the repulsive
interaction from other particles. We suppose the interaction is
on-site only and we tune its strength, in order to interpolate between
the non-interacting and hard-core cases. The $j$-th particle feels the
potential from all remaining ones. To formalize it, we denote $n_j(x,\tau)=\sum_{i=1}^N\overline{\delta}(i-j)\delta(x-x_{i}(\tau))$ 
the number of particles, except
$j$-th particle,  at site
$x$. (We use $\delta(a-b)$ for Kronecker delta and
$\overline{\delta}(a-b)=1-\delta(a-b)$.)
 Thus, the $j$-th
 particle moves 
in the potential
\begin{equation}
U_j(x,\tau)=V(x,\tau)+x\,F+\frac{g}{1-g}\,n_j(x,\tau)\;.
\label{eq:potential}
\end{equation}
For $g=0$ we recover the non-interacting case, while when 
$g\to 1$ we approach the hard-core interaction of the exclusion
process \cite{bly_eva_07}. In contrast with the previous work 
\cite{slanina_08}, we use here 
different form of the interaction in order to see the limit of hard-core repulsion when $g\to 1$.
 Although it may cause some small difficulties when
comparing the results of \cite{slanina_08} with the present ones, the
advantage lies in the possibility to see the transition from
non-interacting case to hard-core repulsion on a finite interval
$g\in[0,1]$.

The simulation algorithm proceeds as follows. At each integer time
$\tau$ we instantly shift the potential according to
(\ref{eq:time-dep-potential}). Then, we choose 
$N$ times a particle randomly and let it try to
make a jump. Therefore, on average every particle makes one attempt per
one time unit, but the probability that a given particle performs
actually $k$ attempts approaches Poisson distribution with unit mean,
$P(k)=1/(\mathrm{e}\,k!)$, when $N$ is large. For small $N$ there is a
deviation from the Poisson distribution, which induces slight
finite-size effects, but in \cite{slanina_08} we showed that they can
be neglected for $N$ larger than about $100$. 
Note that in each of these $N$ attempts the external potential $V(x,\tau)$ is
the same, but the potential $U_j(x,\tau)$ felt by the particle $j$ may
change, because the configuration of particles $n_j(x,\tau)$ changes
after each particle jump.

In one attempt, the particle is allowed to jump one site to the right
or left. The probability of the jump $x\to y$ of the $j$-th particle is
\begin{equation}
W_{j,x\to y}=\frac{1}{2}\,
\Big(1+\mathrm{e}^{\beta\,(U_j(y,\tau)-U_j(x,\tau))}\Big)^{-1}
\label{eq:hopping-probability}
\end{equation}
if $|x-y|=1$ and zero if $|x-y|>1$. 
For convenience, we define the temperature $T$ so that $\beta=270/T$.

Let us now specify the main measurables. 
The simplest quantity of interest is the current
\begin{equation}
J(\tau)=\sum_{i}\big(x_i(\tau+1)-x_i(\tau)\big)
\end{equation}
or rather its time average per particle
$J=\lim_{\tau\to\infty}(\tau\,N)^{-1}\,\sum_{\tau'=1}^\tau J(\tau')$. 
As we are interested in the  energetics of the motor, we must
define the energy input and the useful work done by the particle. 
The latter quantity, at
time $\tau$, is simply $w(\tau)=F\,J(\tau)$. The
 energy pumped into the
motor from outside relates to the change of the potential
$V_a(\tau)$  while the particles stay immobile. Thus, the energy
absorbed by the particle $i$
between steps $\tau - 1$ and $\tau$ is
\begin{equation}
a_i(\tau) = V(x_i(\tau),\tau)-V(x_i(\tau),\tau-1)
\end{equation}
and the efficiency, accordingly, 
\begin{equation}
\eta=\lim_{\tau\to\infty}
\frac{\sum_{\tau'=1}^\tau
  w(\tau')
}{
\sum_{\tau'=1}^\tau\sum_i a_i(\tau')}\;.
\end{equation}
Later in this paper we shall investigate the distribution of the
particle shift 
\begin{equation}
P(\Delta x) =\frac{1}{\cal{N}}\sum_\tau\sum_i\delta\Big(x_i(\tau+\Delta
\tau)-x_i(\tau)-\Delta x \Big)
\end{equation}
and also the joint distribution with the input energy
\begin{equation}
\begin{split}
P&(E_\mathrm{in},\Delta x)=\\ 
&=\frac{1}{\cal{N}}\sum_\tau\sum_i
\delta\Big(\sum_{\tau'=\tau+1}^{\tau+\Delta\tau}a_i(\tau')-E_\mathrm{in}\Big)
\times\\
&\times
\delta\Big(x_i(\tau+\Delta
\tau)-x_i(\tau)-\Delta x \Big)\end{split}
\label{eq:def-distr-ein-dx}
\end{equation}
where $\cal{N}$ is the appropriate normalization. Note that in both
 distributions there is implicit dependence on the time lag $\Delta \tau$.

\begin{figure}[t]
\includegraphics[scale=0.85]{%
\slaninafigdir/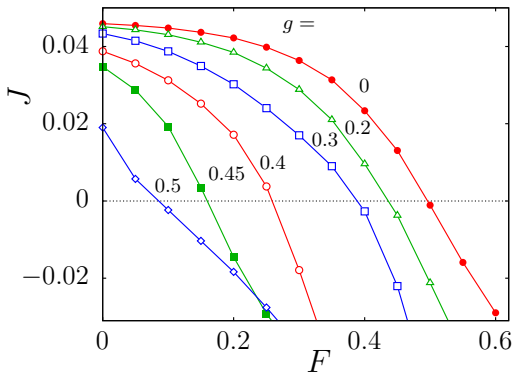}
\caption{%cur-vs-force-1200-600-10-1-16-all-x...
(Color online) 
Dependence of the current on the load, for $L=1200$, $N=600$, $T=10$,
$t=16$
and interaction $g=0$ ({\Large $\bullet$}), $0.2$ ($\bigtriangleup$), $0.3$
($\Box$), $0.4$ ({\Large $\circ$}), $0.45$ ($\blacksquare$), $0.5$
({\Large $\diamond$}). 
}
\label{fig:cur-vs-force-1200-600-10-1-16-all-x}
\end{figure}

\section{Enhanced efficiency}

We show in Fig. \ref{fig:spacetime} examples of typical evolutions
of the particle configurations, for three densities ($\rho=1/2$,
$\rho=2/3$, and $\rho=1$) and three interaction strengths ($g=0$,
$g=0.5$, and $g=0.9$). We can see that without interaction, particles
are bunched together and dragged by the traveling wave of the
periodic potential. Interaction smears out this picture, suppresses the
current and makes at the
same time the local particle density more uniform.

The typical dependence of the current and efficiency on the
interaction strength is shown in
Fig. \ref{fig:eff-vs-int-1200-1200-all-16-x-1}. At zero temperature,
the dependence contains many steep steps with multiple maxima and
minima. Therefore, for some values of the external load $F$ the
current changes sign several times when the interaction increases. For
larger temperatures there are still visible traces of this complex
dependence, although the singularities (sharp steps) are smeared out. 
We also observe that both the current and efficiency approaches zero
for very strong repulsion ($g\to 1$). We shall see later that this
feature is special to some values of the particle density
$\rho=N/L$, for example to $\rho=1$, which applies to
Fig. \ref{fig:eff-vs-int-1200-1200-all-16-x-1}. The generic feature is
that for interaction above about 
$g\simeq 0.6$ the current and efficiency approach a constant value. 

The most important finding, from the point of view of practical use of
the motors, is the increase of the efficiency when the interaction is
switched on but is not too strong. For zero temperature we observe
multiple maxima of the 
efficiency, which transform into a unique maximum at higher
temperatures. The effect of efficiency enhancement was investigated in
detail in our previous work \cite{slanina_08}. In this paper we return
to the origin of this effect later, when we shall discuss the energy
balance and work
fluctuations. 

In Fig. \ref{fig:cur-vs-int-1200-360-10-16-x-all} we can see three
sets of results for the current, differing in the density of
particles. Different curves in one set correspond to different
external load 
$F$. Al three cases (and also the data shown in
Fig. \ref{fig:eff-vs-int-1200-1200-all-16-x-1}) exhibit minimum
current, i. e. smallest effective driving, at interactions somewhere
around $g\simeq 0.4$ to $g\simeq 0.5$. In order to see what is special in this
value of the interaction, we should note that the change of the
potential due to presence of a single particle $g/(1-g)$ is equal to
the amplitude of the traveling-wave potential $V=1$ just for
$g=0.5$. At this value of the interaction, one particle may block, or
at least significantly hinder, the movement of the remaining
particles.

\begin{figure}[t]
\includegraphics[scale=0.85]{%
\slaninafigdir/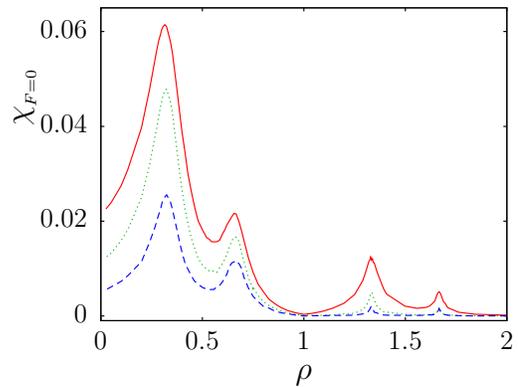}
\caption{%response-x-1200-all-16-9...
(Color online) 
Dependence of the response at zero load on particle density, for 
 $N=1200$,  $t=16$, $g=0.9$ and the temperature $T=30$ (solid
line), $T=10$ (dotted line), and $T=0$ (dashed line).}
\label{fig:response-x-1200-all-16-9}
\end{figure}

We can see that for low density, $\rho<0.5$, the asymptotic current
for strongly interacting particles, $g\to 1$, is positive at low load
and at the same time is sensitive to the value of the load. On the
other hand, for $0.5<\rho<1$ the asymptotic current at zero load is
negative, i. e, 
the interaction induces current reversal. Contrary to the previous
case, the asymptotic current seems to be extremely weakly  dependent
on the load. The third panel shows again that the asymptotic current
is zero for unit density, independently of the load. 

Complementary information can be read from
Fig. \ref{fig:cur-vs-force-1200-600-10-1-16-all-x}, showing the
dependence of the current on the load. We can observe, how the current
decreases with the interaction in the full range of observed $F$. As a
consequence, also the stopping force, i. e. the value of $F$ for which
$J=0$, decreases with increasing interaction. It is also interesting to
note the non-linear decrease of the current with the load. So, the
response function, defined as the derivative $\mathrm{d}J/\mathrm{d}F$,
depends on $F$.

\begin{figure}[t]
\includegraphics[scale=0.85]{%
\slaninafigdir/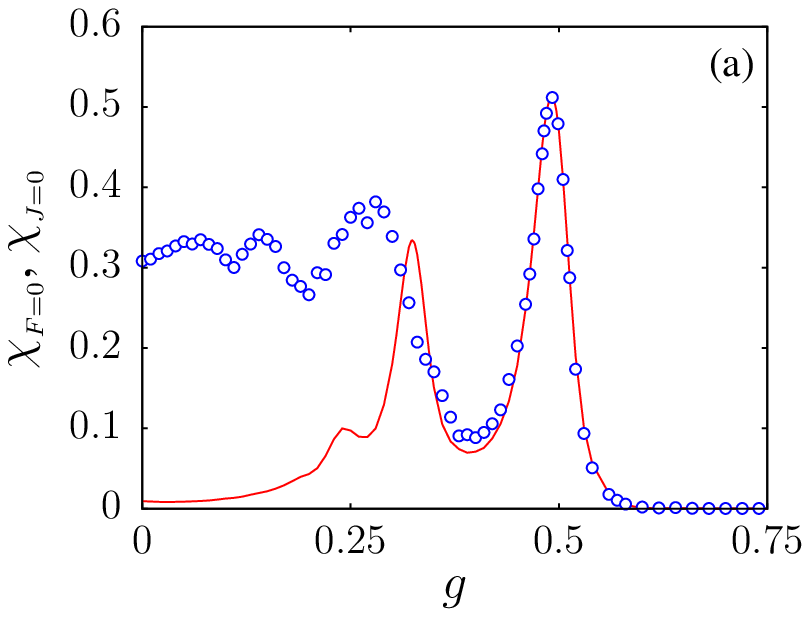}
\includegraphics[scale=0.85]{%
\slaninafigdir/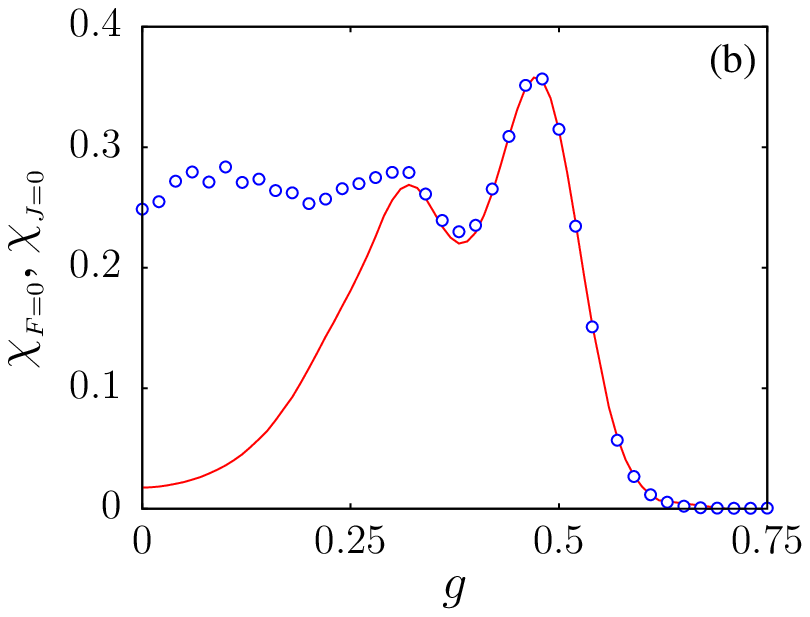}
\caption{%response-zerof-zeroj-1200-1200-10-16-x...
(Color online) 
Dependence of the response on interaction strength, for 
 $L=N=1200$, and $t=16$. Lines: response at zero load; symbols:
response at zero current. The temperature is
 $T=10$ (panel (a)), $T=30$
(panel (b)). 
}
\label{fig:response-zerof-zeroj-1200-1200-10-16-x}
\end{figure}

In Fig. \ref{fig:response-x-1200-all-16-9} we can see how the
zero-load response function 
\begin{equation}
\chi_{_{F=0}}=-\lim_{F\to 0}\frac{\mathrm{d}J}{\mathrm{d}F}
\end{equation}
depends on the density, in the regime of  very strong but finite
repulsion ($g=0.9$).  
Globally, the response is stronger at higher temperature, which is due
to the fact that  at low temperature the movement of the particles is
determined to larger extent by the traveling wave, with lesser
influence of the external load, provided the load is small. 
An interesting feature is the structure of the peaks and the minima seen in 
 Fig. \ref{fig:response-x-1200-all-16-9} at all temperatures. At
 integer values of the density the response approaches  zero. The
 other minima are not so deep and are located at densities slightly
 above the values 
 $\rho=1/2$, $\rho=3/2$ etc. Interestingly, the maxima are found at
 densities very close to the fractions $\rho=1/3$,  $\rho=2/3$,
 $\rho=4/3$, and  $\rho=5/3$.  

As we already said, the response depends on the load, so we must
distinguish from $\chi_{_{F=0}}$ at least one more response function,
defined at zero current
\begin{equation}
\chi_{_{J=0}}=-\Big(\lim_{J\to 0}\frac{\mathrm{d}F}{\mathrm{d}J}\Big)^{-1}\;.
\end{equation}
We can compare these two quantities in
Fig. \ref{fig:response-zerof-zeroj-1200-1200-10-16-x}.  The difference
between 
$\chi_{_{F=0}}$ and $\chi_{_{J=0}}$ is especially marked for low
interaction strength, while at about $g\simeq 0.3$ they come close to
each other and at $g\simeq 0.5$ the two become
nearly indistinguishable. The source of this behavior can be
understood looking at
Fig. \ref{fig:cur-vs-force-1200-600-10-1-16-all-x}. 
Without interaction, the
dependence of the current on the external load is markedly non-linear,
so that the derivative at $F=0$ and $J=0$ differ. Increasing the
interaction, the non-linearity weakens and at $g\simeq 0.5$ we observe
nearly linear dependence, resulting in nearly equal values of the
derivative at  $F=0$ and $J=0$.   Note that the
density is $\rho=1$ in
Fig. \ref{fig:response-zerof-zeroj-1200-1200-10-16-x}
 and both response functions
approach zero when $g\to 1$, in accordance with the results shown in
Fig.  \ref{fig:response-x-1200-all-16-9}. 

It is interesting that the
dependence on $g$ exhibits several peaks. The last (and highest) one
is located close to $g=1/2$ and has nearly the same shape both in
$\chi_{_{F=0}}$ and $\chi_{_{J=0}}$. However, at lower $g$ the peaks 
in the two response functions are much different. We already mentioned
that the interaction $g=1/2$ is special, as the change in potential
due to presence of a single particle just equals the amplitude of the
periodic potential $V(x)$. Also the second highest peak in $\chi_{_{F=0}}$
seems to be located at a special value of the interaction strength,
namely close to $g=1/3$. We can also see a small peak close to
$g=1/4$. We believe these special values are due to special blocking
configurations of particles, which enhance the sensitivity of the
system to the presence of the external load. Indeed, $g=1/3$ means
that two particles on the same site contribute as much as the
amplitude of  $V(x)$, at $g=1/4$ the same holds for three particles
at a site. 

To avoid confusion, we do not claim that the configurations
of one, two, three, etc. particles are more (or less) frequent
at certain values of $g$. What we claim is the following. These
configurations happen time to time. When they do happen,
and if $g$ has special values, they  cause large sensitivity to the
value of the load. For other
 values of $g$, the
sensitivity to the load is weaker, whatever configuration of particles occurs.

\begin{figure}[t]
\includegraphics[scale=0.85]{%
\slaninafigdir/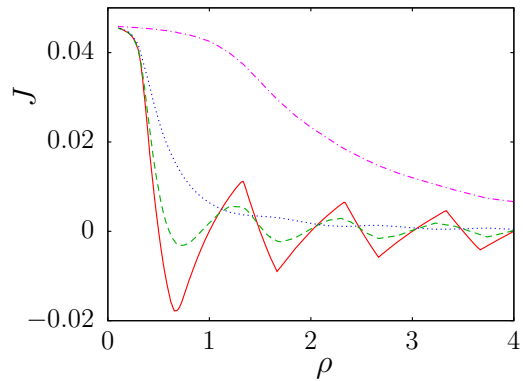}
\caption{%cur-vs-density-10-1-16-all-0...
(Color online) 
Dependence of the current on particle density, for $N=1200$, $T=10$, $t=16$,
$F=0$ and interaction $g=0.9$ (solid line), $g=0.52$ (dashed line),
$g=0.49$ (dotted line), and $g=0.3$ (dash-dotted line).
}
\label{fig:cur-vs-density-10-1-16-all-0}
\end{figure}

\begin{figure}[t]
\includegraphics[scale=0.85]{%
\slaninafigdir/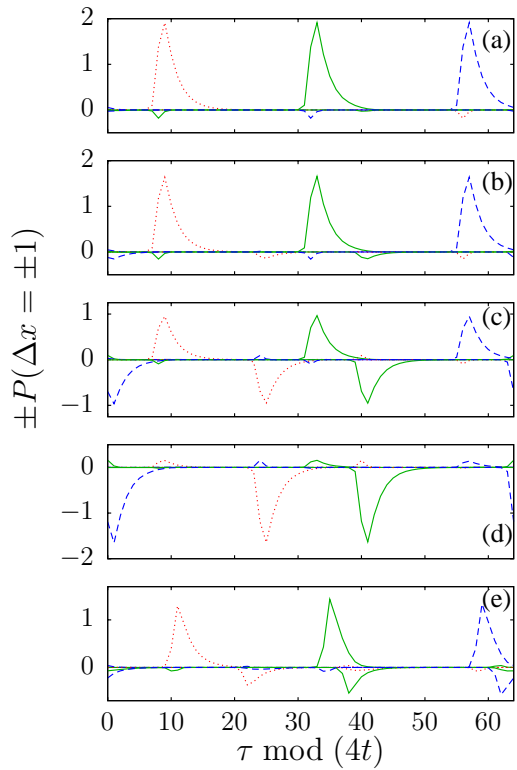}
\caption{%count-plus-minus-shift-x-10-16-x-0...
(Color online) 
The weight of forward (positive quantities) and minus weight of
backward (negative quantities) steps performed at instants
$\tau'=\tau\;\mathrm{mod}\;(4t)$ within the period. The steps
originate at points $x'=x\;\mathrm{mod}\;3$, where $x'=0$ corresponds
to dotted line, $x'=1$ to solid line, and $x'=2$ to dashed line. The
five panels have the following parameters, starting from the
top.
(a): $N=400$, $g=0$, $F=0$. (b):  $N=400$, $g=0.9$, $F=0$. 
(c): $N=600$, $g=0.9$, $F=0$. (d):  $N=800$, $g=0.9$, $F=0$.
 (e):  $N=400$, $g=0.9$, $F=0.3$. 
 All five: $L=1200$, $T=10$, $t=16$. }
\label{fig:count-plus-minus-shift-x-10-16-x-0}
\end{figure}

We also looked at the density dependence of the current at high
density and strong interaction. The results are summarized in
Fig. \ref{fig:cur-vs-density-10-1-16-all-0}. For the strongest
interaction investigated, $g=0.9$, the curve $J(\rho)$ has a  very
peculiar zig-zag shape, with zeros at $\rho=m/2$, maxima at $\rho=m+1/3$,
and minima at $\rho=m-1/3$, for positive integer $m$. When the interaction
is weakened, the sharp cusps become mild waves, until the structure of
maxima and minima vanishes at about $g=0.5$. For smaller $g$, the
current is a monotonously decreasing function of density. 

Note that the motor with hard-core repulsion undergoes a current reversal at a
density within the interval $\rho\in(0,1)$. This is in sharp contrast
with the asymmetric exclusion process, where the current is
proportional to $\rho(1-\rho)$. The reason for this difference lies in
rather different way the particles are driven. In ASEP, there is
constant and homogeneous drift, only hindered by the exclusion
principle. In our model, the driving originates from the time
dependence of the potential, therefore, it is also space- and
time-dependent. The orientation of the current depends on precise
timing of the potential changes at different places. The interaction
changes the potential a particle feels and the current is susceptible
to the details of the potential, so there is no guarantee that the
orientation of the current will be the same with interaction as it
was without interaction. Indeed, the current reversal phenomenon
demonstrates that the orientation of the current does change due to
the interaction. Note also that the current reversal was observed
(qualitatively correctly) also in the approximate mean-field
calculation \cite{slanina_09}.

Some insight into the current reversal phenomenon can be gained from
the statistics of forward and backward steps at different places and
different times within the period. We define the measured weight 
$ P(\Delta x=\pm 1;x',\tau')$
as the average number of particles which jump forward (``$+$'' sign) and
backward (``$-$'' sign) from site $x$ at time $\tau$, where $x'=x\mod 3$ and
$\tau'=\tau\mod (4t)$. Note that it is not a probability, because it
is not normalized to unity. We can see a typical example in
Fig. \ref{fig:count-plus-minus-shift-x-10-16-x-0}. 
 We can see that without
interaction the particles alternately prefer to jump forward from
sites $x'=0$, $1$, and $2$. The backward jumps are rare. This behavior
is independent of the particle density by definition. If we add
strong repulsion, $g=0.9$, the picture differs substantially in the
low and high density regime. For $\rho=1/3$ the statistics of forward
jumps differs little from the non-interacting case, and the frequency
of backward jumps is increased, but remains low. At half filling,
$\rho=1/2$, the particles jump alternately forward and backward, at
different times, so that the total effect is zero current, as seen
already in Fig. \ref{fig:cur-vs-density-10-1-16-all-0}. When the
density is further increased to $\rho=2/3$, the statistics is nearly a
mirror image of the case $\rho=1/3$. The particles preferably jump
backward at specific places and times, and the forward jumps are
rare. For comparison, we show in the last panel of Fig.
\ref{fig:count-plus-minus-shift-x-10-16-x-0} how the statistics is
influenced by non-zero external load.  The time dependence looks
similar, but weight of forward jumps is suppressed and the weight of
backward ones is enhanced.

As the probability of the jumps reflects the 
local potential,
 and therefore the local instantaneous configuration of particles, 
through the formula (\ref{eq:hopping-probability}),
the statistics of the jumps shown in Fig.
\ref{fig:count-plus-minus-shift-x-10-16-x-0} tells us, what is, on
average, the local neighborhood of a particle at positions $x'$ and
times $\tau'$. Change in the shape of the jump statistics reflects the
reorganization of the local particle configurations due to repulsive
interaction. We can clearly see that the reorganization of the
particles can be so dramatic that the current changes sign. 

For comparison, we show also the statistics of jumps in the presence
of non-zero external load. The suppression of positive and enhancement
of negative peaks is visible, as expected.

\begin{figure}[t]
\includegraphics[scale=0.85]{%
\slaninafigdir/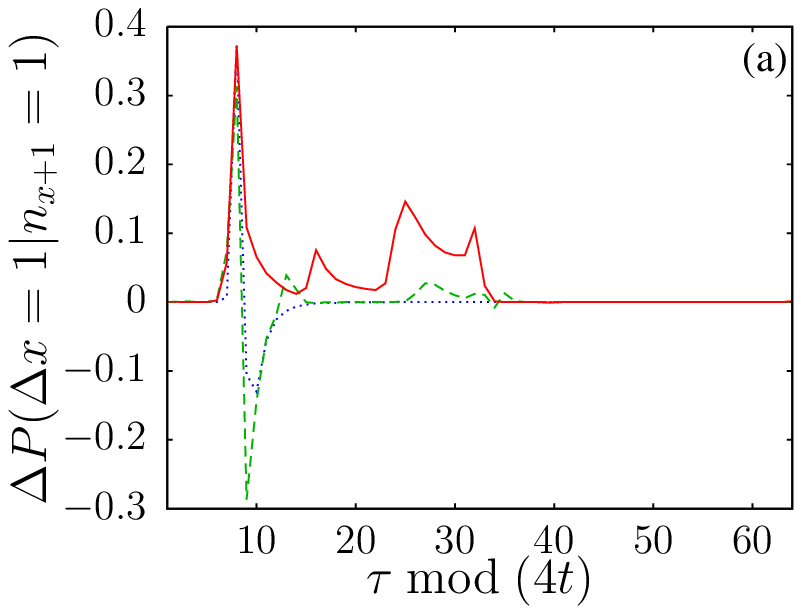}
\includegraphics[scale=0.85]{%
\slaninafigdir/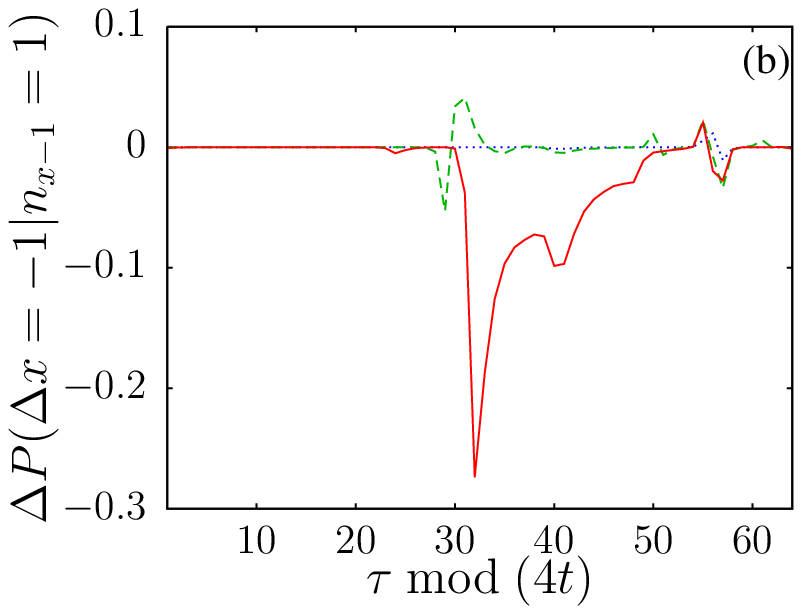}
\caption{%diff-count-minus-to-one-1200-1200-10-1-16-all...
(Color online) 
Sensitivity of the weight of forward (a) and backward (b) steps to the
change in external load. In this statistics, steps originate at site $x$, where
$0=x\mod 3$ and particles go to the site already occupied by exactly one
particle.
 Solid line corresponds to interaction $g=0.5$, dashed line
$g=0.4$, dotted line $g=0$.
Other parameters are 
$N=L=1200$, $T=10$, $t=16$. As for the external load $F$, see the
definition of the plotted quantities in the main text. 
}
\label{fig:diff-count-minus-to-one-1200-1200-10-1-16-all}
\end{figure}

Similar analysis can also make more clear the argument stated before,
that the peaks in the response function at special values of $g$ are
related to the enhanced sensitivity of certain configurations of particles to
external load. For example, for $g=0.5$ such sensitive situation
occurs when a particle tries to hop to a site where there is already a
single particle. To support this view we plot a similar statistics as
in Fig. \ref{fig:count-plus-minus-shift-x-10-16-x-0}, but for the
difference in the count for force $F=0.01$ and opposite $F=-0.01$, on
condition that the site to which the particle is moving, already contains
exactly one other particle. We can write that quantity as
\begin{equation}
\begin{split}
\Delta P(&\Delta x=\pm 1|n_{x\pm 1}=1)\equiv\\
&\equiv
P(\Delta x=\pm 1|n_{x\pm 1}=1)\big|_{F=-0.01}-\\
&-P(\Delta x=\pm 1|n_{x\pm 1}=1)\big|_{F=0.01}
\end{split}
\end{equation}
where $x$ is the original position of the particle, $x\pm 1$ the
position after the move, $n_{x\pm 1}$ number of other particles at the
site where the particle is about to move. We plot an example of this
statistics in
Fig. \ref{fig:diff-count-minus-to-one-1200-1200-10-1-16-all}. 
We compare the situation at interaction $g=0$, $g=0.4$, and
$g=0.5$. We can see that the case $g=0.5$ is indeed special. The
sensitivity to the external load is larger. Moreover, the difference
in statistics has the same sign for almost all instants within the
time period (positive for forward moves, negative for backward ones),
while both for $g=0$ and $g=0.4$ there are positive as well as
negative differences.

\begin{figure}[t]
\includegraphics[scale=0.85]{%
\slaninafigdir/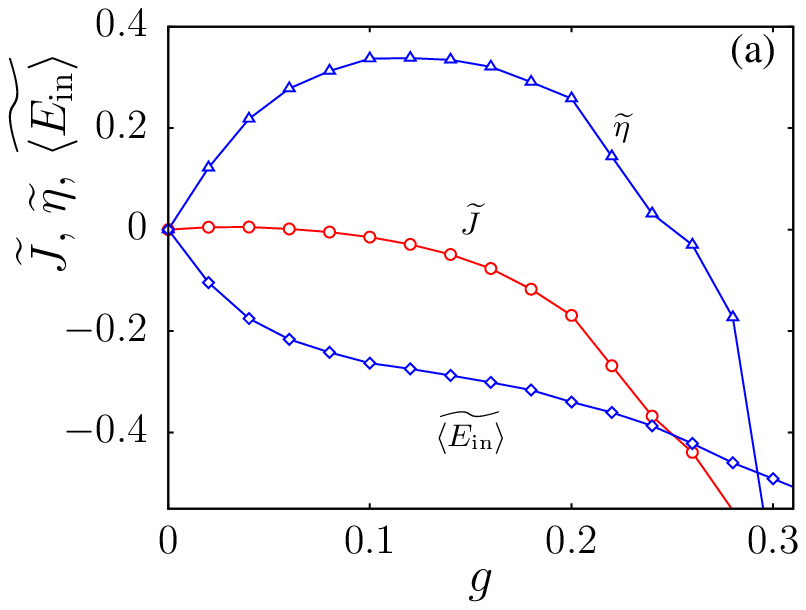}
\includegraphics[scale=0.85]{%
\slaninafigdir/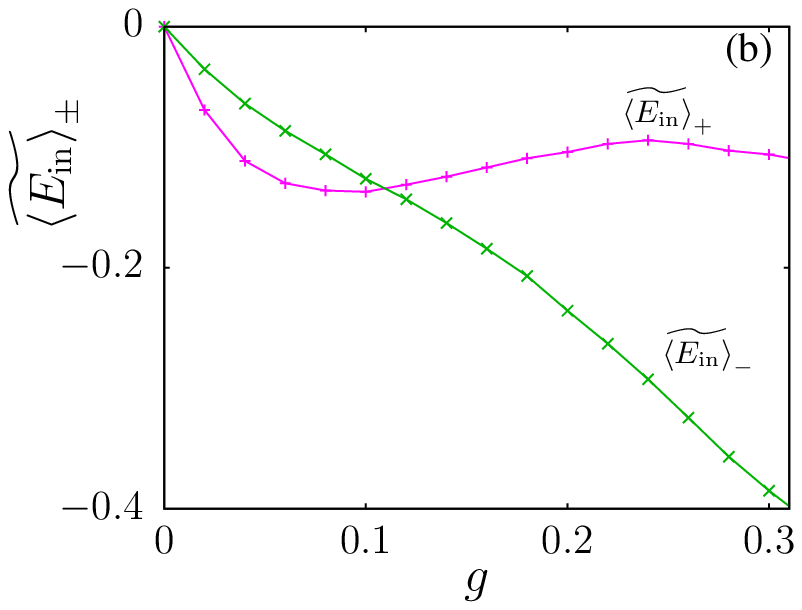}
\caption{%eff-curr-ener-ener-vs-int-10-16-x-1...
(Color online) 
Current ({\Large $\circ$}) , efficiency ($\bigtriangleup$), average
energy input ({\Large $\diamond$}) and its positive ($+$) and
negative ($\times$) parts, all relative to the value at $g=0$. The other
parameters are 
$N=L=1200$, $T=10$, $t=16$, and $F=0.1$. 
}
\label{fig:eff-curr-ener-ener-vs-int-10-16-x-1a}
\end{figure}

\section{Work fluctuations}

To understand better the effect of enhanced efficiency due to
interaction, we shall look at the energy balance and
fluctuations. First, we compare the values of current, efficiency and
average input energy $\langle E_\mathrm{in}\rangle$ relative to their
values at zero interaction, denoted $J_0$, $\eta_0$, and  $\langle
E_\mathrm{in}\rangle_0$, respectively. More precisely, we plot  the typical
interaction dependence of the quantities $\widetilde{J}=J/J_0-1$,
$\widetilde{\eta}=\eta/\eta_0-1$ and $\widetilde{\langle
  E_\mathrm{in}\rangle}=\langle E_\mathrm{in}\rangle/\langle 
E_\mathrm{in}\rangle_0-1$ in
Fig. \ref{fig:eff-curr-ener-ener-vs-int-10-16-x-1a}. We can clearly see
that the initial increase of efficiency for small $g$ is accompanied
by nearly no change in the current, while the input energy
decreases. Therefore, the enhanced efficiency is due to lower energy
input, while the output (the work) remains nearly unchanged. 
When the interaction strength increases further, the
current  starts decreasing as well and so does the work, which is proportional to $J$.
 This effect finally 
outweighs the lower energy input and the efficiency decreases
again. This is the source of the maximum in the efficiency at moderate
values of the interaction.
 
We can get a bit more detailed information if we split the input
energy into its positive and negative parts. Recall, that according to
the definition (\ref{eq:def-distr-ein-dx}) the input energy is
$E_\mathrm{in}=\sum_{\tau'=\tau+1}^{\tau+\Delta\tau}a_i(\tau')$. We
separate the contributions from times when $a_i(\tau')$ is positive
from those when it is negative. The former correspond to the shift of
the potential $V(x,\tau)$ upward, that latter to its downward
move. So, $E_{\mathrm{in}\pm}=\sum_{\tau'=\tau+1}^{\tau+\Delta\tau}\pm
a_i(\tau')\theta\big(\pm a_i(\tau')\big)$, where $\theta(x)$ is the
Heaviside function. With this definition we have $E_\mathrm{in}=
E_{\mathrm{in}+}-E_{\mathrm{in}-}$.  We then define the contributions
from positive and negative potential moves to the quantity 
$\widetilde{\langle   E_\mathrm{in}\rangle}$ as 
$\widetilde{\langle   E_\mathrm{in}\rangle}_+=
(\langle E_{\mathrm{in}+}\rangle-\langle E_{\mathrm{in}+}\rangle_0)/\langle 
E_\mathrm{in}\rangle_0$ and
$\widetilde{\langle   E_\mathrm{in}\rangle}_-=
(\langle E_{\mathrm{in}-}\rangle_0-\langle E_{\mathrm{in}-}\rangle)/\langle 
E_\mathrm{in}\rangle_0$
where, as above, the subscript $0$ denotes the quantities computed at
$g=0$. Hence $\widetilde{\langle
  E_\mathrm{in}\rangle}=\widetilde{\langle
  E_\mathrm{in}\rangle}_++\widetilde{\langle
  E_\mathrm{in}\rangle}_-$. We show the dependence of
$\widetilde{\langle   E_\mathrm{in}\rangle}_\pm$ again in Fig.
\ref{fig:eff-curr-ener-ener-vs-int-10-16-x-1a}.  We can see that both
positive and negative parts contribute to the decrease of the input
energy. The contribution of the positive
part is larger in the most interesting region of moderate $g$, where the
efficiency grows with interaction, while for larger $g$ the decrease
of the negative part becomes more important. This leads to the following
explanation of the 
effect of increased efficiency. 

\begin{figure}[t]
\includegraphics[scale=0.85]{%
\slaninafigdir/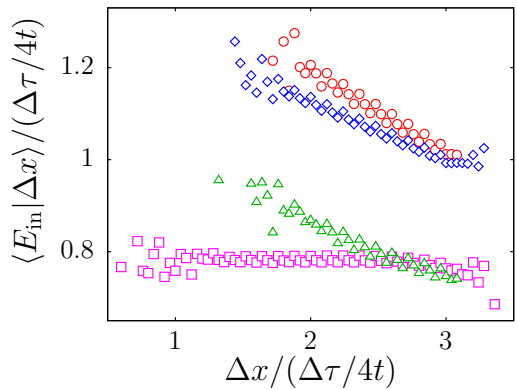}
\caption{%energyin-1200-1200-all-16-all-1a...
(Color online) 
Dependence of the average energy input gained by one particle 
on the shift of the particle. The quarter-period is $t=16$ and the
time lag $\Delta \tau=100 t$ (i. e. $25$ periods). The temperature is
$T=10$ ({\Large $\circ$} and $\bigtriangleup$), $T=30$  ($\Box $ and
{\Large $\diamond$}). The interaction is $g=0$ 
({\Large $\circ$} and {\Large $\diamond$}),
$g=0.1$ ($\Box $  and $\bigtriangleup$). The remaining parameters are
$N=L=1200$, $F=0.1$. The horizontal axis is normalized by
the number of time periods, which is $\Delta\tau/4t$, therefore it
expresses the shift per period.
}
\label{fig:energyin-1200-1200-all-16-all-1a}
\end{figure}

At not too high temperature, the
particles are chiefly driven by the traveling wave of the periodic
potential. This is the power-stroke mechanism of the molecular
motor. When the interaction is switched on, but remains small, the
particles move in an effective potential which differ little from the
original traveling wave. So, the current remains nearly the same, as
testified in Fig. \ref{fig:eff-curr-ener-ener-vs-int-10-16-x-1a}, 
while the input energy is lowered, as is also seen in 
Fig. \ref{fig:eff-curr-ener-ener-vs-int-10-16-x-1a}. 
This lowering could be understood as follows. On
the other hand, the repulsion affects the distribution of particles
within the period of the potential $V(x)$. The minima become
shallower, therefore the particles are less concentrated at them. But
it is the minimum of the potential which is shifted above in the time
evolution of the potential, so it is the particle at the minimum that
acquires the energy from the source of the driving. Less particles at
the minimum equals less input of energy, more precisely lowering of
the positive part of the input energy. Conversely, the particles
pushed off the instantaneous minima of the potential are found at the
maxima of the potential. But these particles suffer lowering of the
time-dependent potential, i. e. returning the energy back to the
external source, therefore lowering also the negative part of the
input energy.   These two
effects, i. e. unchanged current and lowered energy input, are the
explanation of the increased efficiency. Of course, more subtle
effects are also at work here. Especially, also the negative part of
the energy input contributes. More importantly, if the interaction
is strong enough, it changes the potential the particles move in to
such extent that the current diminishes. At very small temperature,
the current is sensitive to tiny changes in the shape of the potential
and small changes in the interaction strength can cause big jumps in
the current. We have seen these jumps in
Fig. \ref{fig:eff-vs-int-1200-1200-all-16-x-1}.

In addition to the averages, we measured also the full joint
distribution function of particle shift and input energy
(\ref{eq:def-distr-ein-dx}). Because the work performed by one
particle  is proportional  to its shift, we have in fact the joint
distribution of performed work and input energy. As a first piece of
information we plot in Fig. \ref{fig:energyin-1200-1200-all-16-all-1a}
the average energy input at fixed value of the particle shift, during
the time interval $\Delta \tau$. We can observe the
already discussed fact that interaction decreases the energy
input. Here we can see that it holds also for most values of the shift,
i. e. work performed by one single particle, separately.

\begin{figure}[t]
\includegraphics[scale=0.85]{%
\slaninafigdir/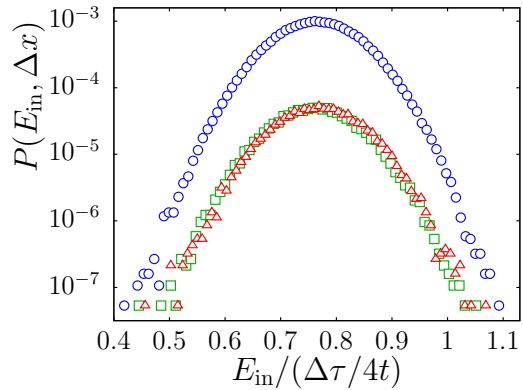}
\caption{%enerdistr-1200-1200-10-16-all-1...
(Color online) 
Probability distribution of the input energy gained by a single particle
at fixed value of the shift of this particle, within the time lag
$\Delta\tau=400\,t$ (i. e. $100$ periods). The fixed shift is $\Delta
x=282$ ({\Large $\circ$}),
$\Delta x=283$ ($\Box$), and
$\Delta x=284$ ($\bigtriangleup$).The other
parameters are 
$N=L=1200$, $T=10$, $t=16$, $g=0.1$, and $F=0.1$. 
The horizontal axis is normalized by
the number of time periods, which is $\Delta\tau/4t$, therefore it
expresses the energy input per period.}
\label{fig:enerdistr-1200-1200-10-16-all-1}
\end{figure}

\begin{figure}[t]
\includegraphics[scale=0.85]{%
\slaninafigdir/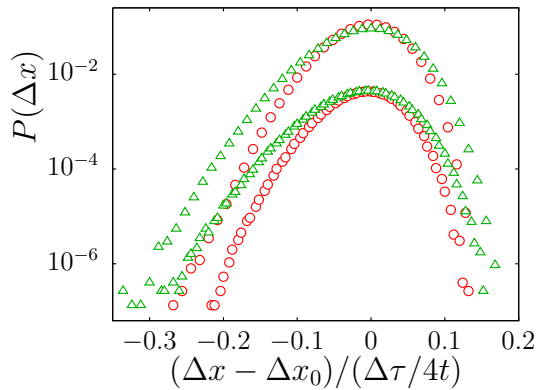}
\caption{%shiftdistr-1200-1200-10-16-all-1...
(Color online) 
Probability distribution of the shift of a particle within the time
lag $\Delta\tau=1000\,t$ (i. e. $250$ periods), 
relative to the most probable value $\Delta x_0$. The interaction
strength is $g=0$ ({\Large $\circ$}, with $\Delta x_0 = 718$) and $g=0.1$
($\bigtriangleup$, with $\Delta x_0 = 718$). The other parameters are
$N=L=1200$, $T=10$, $t=16$, and $F=0.1$. 
The horizontal axis is normalized by
the number of time periods, which is $\Delta\tau/4t$, therefore it
expresses the shift per period.}
\label{fig:shiftdistr-1200-1200-10-16-all-1}
\end{figure}

The probability distribution of the energy input at fixed shift is
shown in Fig. \ref{fig:enerdistr-1200-1200-10-16-all-1}. We can see
that the shape is pretty close to a Gaussian. This is far from being
true for the distribution of the shift, which is proportional to the
work performed by a single particle, as shown in
Fig. \ref{fig:shiftdistr-1200-1200-10-16-all-1}. The distribution is
skewed; when we compare the shifts shorter and longer than the most
probable value, we find that the shorter are significantly more
probable than the longer ones. This is due to the far-from-equilibrium
character of the process.  We can also see that the distribution
is composed of two separate branches. The first one, with higher
probability, corresponds to shifts which are multiples of $3$, the
period of the potential. The other shifts have significantly lower
probability. In fact, it comes as no big surprise, that after integer
number of time periods the particles like to be shifted by integer
number of spatial periods.

The most important finding, however, stems from the comparison of the
distribution in the case with and without interaction. In Fig.
\ref{fig:shiftdistr-1200-1200-10-16-all-1} we make this comparison for
such set of parameters where we know that the interacting case exhibits
higher efficiency. By analogy with equilibrium statistical physics one
might be tempted to guess that higher efficiency is accompanied, or
even caused, by milder fluctuations. The opposite holds in our model
of the molecular motor. The fluctuations of the work performed by
a single particle are higher 
in the interacting case. Therefore, we conclude that the increase of
efficiency is not accompanied by decrease of fluctuations. 
On the contrary, the study
of the energy balance discussed above together with the fact of
increased fluctuations shows that the enhancement of efficiency is
purely an energy effect.

\section{Conclusions}

Interacting molecular motors moving in the power-stroke regime were
modeled using a ``reversible ratchet'' model. Tunable on-site
repulsive interaction leads to a host of intricate phenomena. The most
important of them is the increase of energetic efficiency for small to
moderate values of the interaction strength. We traced the origin of
this effect to energy balance of the process. When the interaction is
increased from zero, the performed work remains practically unchanged,
while the input energy decreases. At the same time, the fluctuations
of the performed work increase. This implies that the enhanced
efficiency does not originate from the suppression of fluctuations,
 contrary to the situation in
equilibrium heat machines. 

Moreover, we  observed that at very low temperature
the dependence of current as well as
efficiency on the interaction strength is rather complex, composed of
many upward and downward steps. Hence, the efficiency has several,
rather than single, local maxima as a function of interaction. As for
the current, for suitable values of the parameters we can observe a
sequence of current reversals when we increase the interaction
strength. This
complicated behavior is due to the interplay between size of steps in
the external periodic potential, in which a particle moves, and the
size of additional 
contributions to the potential from the interaction with other
particles. However, this complicated dependence gradually disappears
when the temperature increases. But the effect of current
reversal due to interaction remains still visible. 

We also investigated the response function of the current with respect
to external load, both for zero load and for zero current. We showed
that these two response functions differ substantially at zero or
small interaction, but become identical when the interaction is
large. We also revealed the structure with several peaks for both
density and interaction dependence of the response function. Detailed
study of the location of these maxima and minima showed that they
correspond to specific fractional values of density and
interaction. For example, the response is zero if the density is
integer number and has maximum for densities equal to integer number
of thirds, except the values which are themselves integers. In the
interaction 
dependence, the peaks were found close to interaction strength equal
to one half, one third and one fourth. We speculate that these special
values are due to the fact that in those cases just one, two, and three
particles on the same site, respectively, contribute to the potential 
by the value exactly equal to the amplitude of the external periodic potential.
Contrary to the
complicated step structure in the current, the peak structure in the
response function survives also at higher temperatures.

The probability distribution of performed work and input energy
reveals that the interaction leads to the increase of fluctuations, as
we already mentioned. But we can see more. First, the distribution of
work is far from Gaussian. It is skewed so that the lower particle
shift (i. e. work performed by an individual particle) 
relative to the maximum is more probable. This is the sign of
far-from-equilibrium regime of the transport in the molecular motor. On the
other hand, the input energy is Gaussian-distributed, when observed at
fixed work. 

There is also a very interesting principal question related to
large-deviation properties of the fluctuation of the performed
work. We made some simulations in this direction, which show that the
work distribution, when properly rescaled, converges to a
large-deviation function. In the last decade, there was a great surge
of activity in the field of Fluctuation Theorems
\cite{eva_coh_mor_93,gal_coh_95,kurchan_98,leb_spo_99,mae_net_03,seifert_05a,and_gas_06,lau_lac_mal_07,astumian_07,har_sch_07,sub_chv_07}
but in our case the problem of applying these results lies in the
choice of the proper quantity which would be  both physically meaningful (or
at least the physical meaning must not be enormously intricate) and
satisfy the Fluctuation Theorem in some of the forms known so
far. This question remains open.

Finally, we must also admit several drawbacks of our model, which can be lifted
only by setting up a completely different scheme of particle
movement. The first point is that the potential changes synchronously at
all sites. This is unrealistic in biological motors, where each
molecule has its internal ``clock'' telling in what phase of the
chemical cycle the motor finds itself. It could be easily possible to
simulate an asynchronous version of the algorithm. On the other hand,
in technological applications the  synchronicity in the potential
changes may be built in into the system. The second point concerns the
tunable interaction used in our model. Motor proteins interact always
by steric hard-core repulsion and th effective weak repulsion may
occur only as projection of real three-dimensional situation onto
one-dimensional effective model \cite{kal_per_05,kal_per_06}. However,
there is no principal 
obstacle to simulate three-dimensional case directly, if only
sufficient computer power is available. Another way out is to
generalize the asymmetric exclusion process in such a way that the
maximum number of particles one
site may accommodate is  not one, but two, or three, etc. Simulations
in this direction are under way.

%As expected, the fluctuation theorem for the particle current does not
%hold, due to the time dependence of the driving. It would be extremely
%interesting to discover the proper quantity for which the fluctuation
%theorem does hold in our model. It could be, a posteriori, identified
%with entropy production. We leave this question open for future
%investigations. 
%
%
%
\begin{acknowledgments}
I gladly acknowledge inspiring discussions with P. Chvosta, E. Ben-Jacob and P. Kalinay. 
This work was carried out within the project AVOZ10100520 of the Academy of 
Sciences of the Czech republic and was 
supported by the Grant Agency of the Czech Republic, grant No. 
202/07/0404.

\end{acknowledgments}

\begin{thebibliography}{99}
%Bib. item no.: 1
%Record: 2542b
\bibitem{schliwa_03}
M. Schliwa (Ed.),
{\it Molecular Motors}
 (Wiley-VCH, New York, 2003).

%Bib. item no.: 2
%Record: 2527
\bibitem{jul_adj_pro_97}
F.\ J\"ulicher, A.\ Ajdari, and J.\ Prost,
Rev. Mod. Phys.
 {\bf 69},
 1269
 (1997).

%Bib. item no.: 3
%Record: 2599
\bibitem{astumian_97}
R.\ D.\ Astumian,
Science
 {\bf 276},
 917
 (1997).

%Bib. item no.: 4
%Record: 2520
\bibitem{rei_han_02}
P.\ Reimann and P.\ H\"anggi,
Appl. Phys. A
 {\bf 75},
 169
 (2002).

%Bib. item no.: 5
%Record: 2528
\bibitem{reimann_02}
P.\ Reimann,
Phys. Rep.
 {\bf 361},
 57
 (2002).

%Bib. item no.: 6
%Record: 2592
\bibitem{sch_woe_03}
M.\ Schliwa and G.\ Woehlke,
Nature
 {\bf 422},
 759
 (2003).

%Bib. item no.: 7
%Record: 2131
\bibitem{han_mar_nor_04}
P. H\"anggi, F. Marchesoni, and F. Nori,
Ann. Phys. (Leipzig)
 {\bf 14},
 51
 (2005).

%Bib. item no.: 8
%Record: 2585
\bibitem{lip_cha_klu_lie_mul_06}
R.\ Lipowsky, Y.\ Chai, S.\ Klumpp, S.\ Liepelt, M.\ J.\ I.\ M\"uller,
Physica A
 {\bf 372},
 34
 (2006).

%Bib. item no.: 9
%Record: 2534
\bibitem{kol_fis_07}
A.\ B.\ Kolomeisky and M.\ E.\ Fisher,
Annu. Rev. Phys. Chem.
 {\bf 58},
 675
 (2007).

%Bib. item no.: 10
%Record: 2605
\bibitem{wan_els_07}
H.\ Wang and T.\ C.\ Elston,
J. Stat. Phys.
 {\bf 128},
 35
 (2007).

%Bib. item no.: 11
%Record: 2594
\bibitem{svo_blo_94}
K.\ Svoboda and S.\ M.\ Block,
Cell
 {\bf 77},
 773
 (1994).

%Bib. item no.: 12
%Record: 2596
\bibitem{wan_ost_98}
H.\ Wang and G.\ Oster,
Nature
 {\bf 396},
 279
 (1998).

%Bib. item no.: 13
%Record: 462
\bibitem{as_bi_96}
R.\ D.\ Astumian and M.\ Bier,
Biophys. J.
 {\bf 70},
 637
 (1996).

%Bib. item no.: 14
%Record: 2597
\bibitem{car_cro_05}
N.\ J.\ Carter and R.\ A.\ Cross,
Nature
 {\bf 435},
 308
 (2005).

%Bib. item no.: 15
%Record: 2595
\bibitem{sch_xia_mer_wei_97}
U.\ Scheer, B.\ Xia, H.\ Merkert, and D.\ Weisenberger,
Chromosoma
 {\bf 105},
 470
 (1997).

%Bib. item no.: 16
%Record: 2525
\bibitem{ras_kob_mal_fid_mas_04}
I.\ Ra\v ska, K.\ Koberna, J.\ Mal\'{\i}nsk\'y, H.\ Fidlerov\'a, and M.\ Ma\v sata,
Biology of the Cell
 {\bf 96},
 579
 (2004).

%Bib. item no.: 17
%Record: 2529
\bibitem{del_benz_slu_aze_gol_06}
P.\ De Los Rios, A.\ Ben-Zvi, O.\ Slutsky, A.\ Azem, and P. Golubinoff,
Proc. Natl. Acad. Sci. USA
 {\bf 103},
 6166
 (2006).

%Bib. item no.: 18
%Record: 2582
\bibitem{han_mar_08}
P.\ H\"anggi and F.\ Marchesoni,
Rev. Mod. Phys.
 {\bf 81},
 387
 (2009).

%Bib. item no.: 19
%Record: 2127
\bibitem{mat_mul_03}
S. Matthias and F. M\"uller,
Nature
 {\bf 424},
 53
 (2003).

%Bib. item no.: 20
%Record: 2123
\bibitem{ket_rei_han_mul_00}
C. Kettner, P. Reimann, P. H\"anggi, and F. M\"uller,
Phys. Rev. E
 {\bf 61},
 312
 (2000).

%Bib. item no.: 21
%Record: 2128
\bibitem{lin_hum_lof_99}
H. Linke, T. E. Humphrey, A. L\"ofgren, A. O. Sushkov, R. Newbury, R. P. Taylor, and P. Omling,
Science
 {\bf 286},
 2314
 (1999).

%Bib. item no.: 22
%Record: 2588
\bibitem{bly_eva_07}
R.\ A.\ Blythe and M.\ R.\ Evans,
J. Phys. A: Math. Theor.
 {\bf 40},
 R333
 (2007).

%Bib. item no.: 23
%Record: 2587
\bibitem{ajd_pro_92}
A. Ajdari and J. Prost,
C. R. Acad. Sci. Paris, S\'erie II
 {\bf 315},
 1635
 (1992).

%Bib. item no.: 24
%Record: 2536
\bibitem{magnasco_93}
M.\ O.\ Magnasco,
Phys. Rev. Lett.
 {\bf 71},
 1477
 (1993).

%Bib. item no.: 25
%Record: 2543
\bibitem{bar_han_kis_94}
R.\ Bartussek, P.\ H\"anggi, and J.\ G.\ Kissner,
Europhys. Lett.
 {\bf 28},
 459
 (1994).

%Bib. item no.: 26
%Record: 2604
\bibitem{kol_wid_98}
A.\ B.\ Kolomeisky and B.\ Widom,
J. Stat. Phys.
 {\bf 93},
 633
 (1998).

%Bib. item no.: 27
%Record: 2602
\bibitem{qian_97}
H.\ Qian,
Biophys. Chem.
 {\bf 67},
 263
 (1997).

%Bib. item no.: 28
%Record: 2603
\bibitem{qian_00a}
H.\ Qian,
Biophys. Chem.
 {\bf 83},
 35
 (2000).

%Bib. item no.: 29
%Record: 2537
\bibitem{qian_04}
H.\ Qian,
Phys. Rev. E
 {\bf 69},
 012901
 (2004).

%Bib. item no.: 30
%Record: 2625
\bibitem{mae_wie_03}
C. Maes and M. H. van Wieren,
J. Stat. Phys.
 {\bf 112},
 329
 (2003).

%Bib. item no.: 31
%Record: 2621
\bibitem{kol_stu_pop_05}
A.\ B.\ Kolomeisky, E.\ B.\ Stukalin, and A.\ A.\ Popov,
Phys. Rev. E
 {\bf 71},
 031902
 (2005).

%Bib. item no.: 32
%Record: 2626
\bibitem{wan_fen_zhe_fan_07}
Z. Wang, M. Feng, W. Zheng, and D. Fan,
Biophys. J.
 {\bf 93},
 3363
 (2007).

%Bib. item no.: 33
%Record: 2589
\bibitem{lip_lie_08}
R.\ Lipowsky and S.\ Liepelt,
J. Stat. Phys.
 {\bf 130},
 39
 (2008).

%Bib. item no.: 34
%Record: 2622
\bibitem{das_kol_08}
R.\ K.\ Das and A.\ B.\ Kolomeisky,
Phys. Rev. E
 {\bf 77},
 061912
 (2008).

%Bib. item no.: 35
%Record: 2627
\bibitem{vilfan_09}
A. Vilfan,
Frontiers in Bioscience
 {\bf 14},
 2269
 (2009).

%Bib. item no.: 36
%Record: 2591
\bibitem{wan_ost_02}
H.\ Wang and G.\ Oster,
Europhys. Lett.
 {\bf 57},
 134
 (2002).

%Bib. item no.: 37
%Record: 2544
\bibitem{jul_pro_95}
F.\ J\"ulicher and J.\ Prost,
Phys. Rev. Lett.
 {\bf 75},
 2618
 (1995).

%Bib. item no.: 38
%Record: 2590
\bibitem{sekimoto_96}
K.\ Sekimoto,
cond-mat/9611005.

%Bib. item no.: 39
%Record: 2538
\bibitem{sekimoto_97}
K.\ Sekimoto,
J. Phys. Soc. Japan
 {\bf 66},
 1234
 (1997).

%Bib. item no.: 40
%Record: 2535
\bibitem{kam_hon_tak_98}
H.\ Kamegawa, T.\ Hondou, and F.\ Takagi,
Phys. Rev. Lett.
 {\bf 80},
 5251
 (1998).

%Bib. item no.: 41
%Record: 2521
\bibitem{par_dec_02}
J.\ M.\ R.\ Parrondo and B.\ J.\ De Cisneros,
Appl. Phys. A
 {\bf 75},
 179
 (2002).

%Bib. item no.: 42
%Record: 2598
\bibitem{asf_bek_05}
M.\ Asfaw and M.\ Bekele,
Phys. Rev. E
 {\bf 72},
 056109
 (2005).

%Bib. item no.: 43
%Record: 2522
\bibitem{par_bla_cao_bri_98}
J.\ M.\ R.\ Parrondo, J.\ M.\ Planco, J.\ F.\ Cao, and R.\ Brito,
Europhys. Lett.
 {\bf 43},
 248
 (1998).

%Bib. item no.: 44
%Record: 2593
\bibitem{wan_ost_02a}
H.\ Wang and G.\ Oster,
Appl. Phys. A
 {\bf 75},
 315
 (2002).

%Bib. item no.: 45
%Record: 2137
\bibitem{parrondo_98}
J. M. R. Parrondo,
Phys. Rev. E
 {\bf 57},
 7297
 (1998).

%Bib. item no.: 46
%Record: 2530
\bibitem{ast_der_99}
R.\ D.\ Astumian and I.\ Der\'enyi,
Biophys. J.
 {\bf 77},
 993
 (1999).

%Bib. item no.: 47
%Record: 2673
\bibitem{par_jul_ajd_pro_99}
A.\ Parmeggiani, F.\ J\"ulicher, A.\ Ajdari, and J.\ Prost,
Phys. Rev. E
 {\bf 60},
 2127
 (1999).

%Bib. item no.: 48
%Record: 2581
\bibitem{sch_sei_08}
T.\ Schmiedl and U.\ Seifert,
Europhys. Lett.
 {\bf 83},
 30005
 (2008).

%Bib. item no.: 49
%Record: 2674
\bibitem{mcd_gib_pip_68}
C.\ T.\ MacDonald, J.\ H.\ Gibbs, and A.\ C.\ Pipkin,
Biopolymers
 {\bf 6},
 1
 (1968).

%Bib. item no.: 50
%Record: 2675
\bibitem{mcd_gib_69}
C.\ T.\ MacDonald and J.\ H.\ Gibbs,
Biopolymers
 {\bf 7},
 707
 (1969).

%Bib. item no.: 51
%Record: 504
\bibitem{do_do_mu_92}
B. Derrida, E. Domany, and D. Mukamel,
J. Stat. Phys.
 {\bf 69},
 667
 (1992).

%Bib. item no.: 52
%Record: 406
\bibitem{de_ev_pa_93}
B. Derrida, M.R. Evans, V. Hakim and V. Pasquier,
J. Phys. A: Math. Gen.
 {\bf 26},
 1493
 (1993).

%Bib. item no.: 53
%Record: 1211
\bibitem{derrida_98}
B. Derrida,
Phys. Rep.
 {\bf 301},
 65
 (1998).

%Bib. item no.: 54
%Record: 2668
\bibitem{par_fra_fre_03}
A. Parmeggiani, T. Franosch, and E. Frey,
Phys. Rev. Lett.
 {\bf 90},
 086601
 (2003).

%Bib. item no.: 55
%Record: 2676
\bibitem{par_fra_fre_04}
A.\ Parmeggiani, T.\ Franosch, and E.\ Frey,
Phys. Rev. E
 {\bf 70},
 046101
 (2004).

%Bib. item no.: 56
%Record: 2532
\bibitem{gre_gar_nis_sch_cho_07}
P.\ Greulich, A.\ Garai, K.\ Nishinari, A.\ Schadschneider, and D.\ Chowdhury,
Phys. Rev. E
 {\bf 75},
 041905
 (2007).

%Bib. item no.: 57
%Record: 2533
\bibitem{bas_cho_07}
A.\ Basu and D.\ Chowdhury,
Phys. Rev. E
 {\bf 75},
 021902
 (2007).

%Bib. item no.: 58
%Record: 2531
\bibitem{tri_cho_08}
T.\ Tripathi and D.\ Chowdhury,
Phys. Rev. E
 {\bf 77},
 011921
 (2008).

%Bib. item no.: 59
%Record: 21
\bibitem{na_sch_92}
K. Nagel and M. Schreckenberg,
J. Phys. I France
 {\bf 2},
 2221
 (1992).

%Bib. item no.: 60
%Record: 2540
\bibitem{lip_klu_nie_01}
R.\ Lipowsky, S.\ Klumpp, and T.\ M.\ Nieuwenhuizen,
Phys. Rev. Lett.
 {\bf 87},
 108101
 (2001).

%Bib. item no.: 61
%Record: 2584
\bibitem{klu_lip_05}
S.\ Klumpp and R.\ Lipowsky,
Proc. Natl. Acad. Sci. USA
 {\bf 102},
 17284
 (2005).

%Bib. item no.: 62
%Record: 2135
\bibitem{der_vic_95}
I. Der\'enyi and T. Vicsek,
Phys. Rev. Lett.
 {\bf 75},
 374
 (1995).

%Bib. item no.: 63
%Record: 2583
\bibitem{der_ajd_96}
I.\ Der\'enyi and A.\ Ajdari,
Phys. Rev E
 {\bf 54},
 R5
 (1996).

%Bib. item no.: 64
%Record: 2136
\bibitem{agh_men_pli_99}
Y. Aghababaie, G. I. Menon, and M. Plischke,
Phys. Rev. E
 {\bf 59},
 2578
 (1999).

%Bib. item no.: 65
%Record: 2523
\bibitem{rei_kaw_bro_han_99}
P.\ Reimann, R.\ Kawai, C.\ van den Broeck, and P.\ H\"anggi,
Europhys. Lett.
 {\bf 45},
 545
 (1999).

%Bib. item no.: 66
%Record: 2624
\bibitem{stu_phi_kol_05}
E.\ B.\ Stukalin, H.\ Phillips III, and A.\ B.\ Kolomeisky,
Phys. Rev. Lett.
 {\bf 94},
 238101
 (2005).

%Bib. item no.: 67
%Record: 2524
\bibitem{stu_kol_06}
E.\ B.\ Stukalin and A.\ B.\ Kolomeisky,
Phys. Rev. E
 {\bf 73},
 031922
 (2006).

%Bib. item no.: 68
%Record: 2601
\bibitem{cqm_kqf_zel_cas_joa_06}
O.\ Camp\`as, Y.\ Kafri, K.\ B.\ Zeldovich, J.\ Casademunt, and J.-F.\ Joanny,
Phys. Rev. Lett.
 {\bf 97},
 038101
 (2006).

%Bib. item no.: 69
%Record: 2600
\bibitem{jul_pro_97}
F.\ J\"ulicher and J.\ Prost,
Phys. Rev. Lett.
 {\bf 78},
 4510
 (1997).

%Bib. item no.: 70
%Record: 2667
\bibitem{bad_jul_pro_02}
M.\ Badoual, F.\ J\"ulicher, and J.\ Prost,
Proc. Natl. Acad. Sci. USA
 {\bf 99},
 6696
 (2002).

%Bib. item no.: 71
%Record: 2623
\bibitem{art_mor_kol_08}
M.\ N.\ Artyomov, A.\ Yu.\ Morozov, and A.\ B.\ Kolomeisky,
Phys. Rev. E
 {\bf 77},
 040901(R)
 (2008).

%Bib. item no.: 72
%Record: 10037
\bibitem{slanina_08}
F. Slanina,
Europhys. Lett.
 {\bf 84},
 50009
 (2008).

%Bib. item no.: 73
%Record: 10038
\bibitem{slanina_09}
F. Slanina,
J. Stat. Phys.
 {\bf 135},
 935
 (2009).

%Bib. item no.: 74
%Record: 2606
\bibitem{eva_coh_mor_93}
D.\ J.\ Evans, E.\ G.\ D.\ Cohen, and G.\ P.\ Morriss,
Phys. Rev. Lett.
 {\bf 71},
 2401
 (1993).

%Bib. item no.: 75
%Record: 2607
\bibitem{gal_coh_95}
G.\ Gallavotti and  E.\ G.\ D.\ Cohen,
Phys. Rev. Lett.
 {\bf 74},
 2694
 (1995).

%Bib. item no.: 76
%Record: 856
\bibitem{kurchan_98}
J.\ Kurchan,
J. Phys. A: Math. Gen.
 {\bf 31},
 3719
 (1998).

%Bib. item no.: 77
%Record: 2610
\bibitem{leb_spo_99}
J.\ L.\ Lebowitz and H.\ Spohn,
J. Stat. Phys.
 {\bf 95},
 333
 (1999).

%Bib. item no.: 78
%Record: 2612
\bibitem{mae_net_03}
C.\ Maes and K.\ Neto\v{c}n\'y,
J. Stat. Phys.
 {\bf 110},
 269
 (2003).

%Bib. item no.: 79
%Record: 2616
\bibitem{seifert_05a}
U.\ Seifert,
Europhys. Lett.
 {\bf 70},
 36
 (2005).

%Bib. item no.: 80
%Record: 2617
\bibitem{and_gas_06}
D.\ Andrieux and P.\ Gaspard,
Phys. Rev. E
 {\bf 74},
 011906
 (2006).

%Bib. item no.: 81
%Record: 2618
\bibitem{lau_lac_mal_07}
A.\ W.\ C.\ Lau, D.\ Lacoste, and K.\ Mallick,
Phys. Rev. Lett.
 {\bf 99},
 158102
 (2007).

%Bib. item no.: 82
%Record: 2620
\bibitem{astumian_07}
R.\ D.\ Astumian,
Phys. Rev. E
 {\bf 76},
 020102(R)
 (2007).

%Bib. item no.: 83
%Record: 2614
\bibitem{har_sch_07}
R.\ J.\ Harris and G.\ M.\ Sch\"utz,
J. Stat. Mech.
 P07020
 (2007).

%Bib. item no.: 84
%Record: 2679x
\bibitem{sub_chv_07}
E.\ \v{S}ubrt and P.\ Chvosta,
J. Stat. Mech.
 P09019
 (2007).

%Bib. item no.: 85
%Record: 2677x
\bibitem{kal_per_05}
P.\ Kalinay and J.\ K.\ Percus,
J. Chem. Phys.
 {\bf 122},
 204701
 (2005).

%Bib. item no.: 86
%Record: 2678x
\bibitem{kal_per_06}
P.\ Kalinay and J.\ K.\ Percus,
Phys. Rev. E
 {\bf 74},
 041203
 (2006).

%
%
\end{thebibliography}
\end{document}